%% file: main_TDSC.tex
\documentclass[lettersize,journal]{IEEEtran}
\usepackage{amsmath,amsfonts}
\usepackage{algorithmic}
\usepackage{algorithm}
\usepackage{array}
\usepackage[caption=true,font=normalsize,labelfont=sf,textfont=sf]{subfig}
\usepackage{textcomp}
\usepackage{stfloats}
\usepackage{url}
\usepackage{verbatim}
\usepackage{graphicx}
\usepackage{enumitem}
\usepackage{multirow} 
\usepackage{cite}
\begin{document}

\title{Deep-Fake CAPTCHA: Mitigating Next-Generation Social Engineering Attacks}

\author{Guy~Frankovits,
        Lior~Yasur,
        Fred~M.~Grabovski,
        and~Yisroel~Mirsky% <-this % stops a space
\thanks{G. Frankovits, L. Yasur, F. M. Grabovski, and Y. Mirsky are with Ben-Gurion University, 1 Ben-Gurion Ave., Beersheba 8410501, Israel (e-mail: guyfrank@post.bgu.ac.il; lioryasu@post.bgu.ac.il; freddie@post.bgu.ac.il; yisroel@bgu.ac.il).}% <-this % stops a space
\thanks{Corresponding author: Yisroel Mirsky (e-mail: yisroel@bgu.ac.il).}}

\markboth{Journal of \LaTeX\ Class Files,~Vol.~14, No.~8, August~2021}%
{Frankovits \MakeLowercase{\textit{et al.}}: Deep-Fake CAPTCHA: Mitigating Next-Generation Social Engineering Attacks}


\maketitle

\begin{abstract}
This paper presents DF-CAPTCHA, an active defense against real-time deepfake impersonation in voice and video calls. Instead of passively searching for artifacts, DF-CAPTCHA prompts the caller to perform simple challenge-response tasks that are easy for humans but difficult for current real-time deepfake systems to generate convincingly. The framework verifies the response using four criteria: realism, identity consistency, task completion, and response time. We evaluate the approach across both audio and video modalities using user studies and experiments with real-time deepfake models. Results show that people often struggle to distinguish real-time deepfakes from authentic media, while DF-CAPTCHA substantially improves detection performance over passive methods, reaching high accuracy in both modalities. These findings suggest that active challenge-based verification is a practical and robust defense against next-generation social engineering attacks based on real-time deepfakes.
\end{abstract}



\section{Introduction}
Deepfakes, defined as media generated by deep neural networks that appear authentic to human perception \cite{mirsky2021creation}, have rapidly evolved since their emergence in 2017. This technology has found applications across various domains, enhancing productivity, revolutionizing education, and providing novel forms of entertainment \cite{VoiceClo10:online,Deepfake15:online,ShamookS49:online}. However, the same technological advancements that enable these positive applications also harbor a dark side, opening avenues for unethical and malicious exploitation.

The ability to convincingly impersonate a target's face and voice has empowered bad actors to conduct a range of nefarious activities. These include defamation campaigns \cite{Melonisu84:online}, propaganda \cite{Deepfake46:online}, blackmail attempts \cite{Deepfake67:online}, the spread of misinformation \cite{Indiasge48:online}, and sophisticated social engineering attacks targeting both individuals and organizations globally \cite{Reshapin92:online}. The misuse of deepfake technology has manifested in various forms, from the creation of non-consensual explicit content \cite{Deepfake76:online} to more insidious political manipulations.

% A particularly alarming example occurred in March 2022 during the Russian-Ukraine conflict, where a deepfake video circulated showing Ukraine's prime minister falsely instructing troops to surrender \cite{Deepfake46:online}. This incident underscores the potential for deepfakes to be weaponized in information warfare, potentially influencing geopolitical events and public opinion on a global scale.

As deepfake technology continues to advance in quality and accessibility, the line between authentic and fabricated media becomes increasingly blurred. This technological progression presents a growing challenge to digital trust and information integrity, necessitating robust detection methods and ethical frameworks to mitigate the risks associated with deepfake misuse.

\subsection{Real-time Deepfakes (RT-DF)}
Recent advancements in deepfake technology have significantly improved its efficiency, giving rise to real-time deepfakes (RT-DFs)\footnote{Examples of RT-DF tools:
\url{https://github.com/iperov/DeepFaceLive}\\
\url{https://github.com/hacksider/Deep-Live-Cam}\\
\url{https://apps.apple.com/us/app/real-deep-real-time-deep-fake/id1639829534}\\
\url{https://www.respeecher.com/}}.
This evolution enables malicious actors to impersonate individuals during live voice and video calls, presenting an unprecedented threat to digital security and trust. The danger of RT-DFs is amplified by three key factors: (1) the unexpected nature of the attack vector, (2) the tendency to mistake familiarity for authenticity, and (3) the rapidly improving quality of RT-DF technologies.

To conceptualize this threat, let's perform the following thought experiment. Imagine someone receives a call from their mother who is in trouble and urgently needs a money transfer. The caller sounds exactly like her, but the situation seems a bit out of place. Under stress and frustration, she hands the phone over to someone who sounds like the victim's father, who confirms the situation. Without hesitation, many would transfer the money even though they're technically talking to a stranger.

Now consider state-actors with considerable amounts of time and resources. They could target workers at power plants and other critical infrastructure by posing as their administrators. Over a phone call, they could convince the worker to change a configuration or reveal confidential information which would lead to a cyber breach or a catastrophic failure. Attackers could even pose as military officials or politicians leading to a breach of national security. 

These scenarios are plausible because some existing real-time frameworks can impersonate an individual's face or voice using very little information. For example, some real-time methods can reenact a face with one sample image \cite{NIPS2019_8935} and some can clone a voice with just a few seconds of audio \cite{han2024vall}. Using these technologies, an attacker would only need to call the source voice for a few seconds or scrape the source's image from the internet to perform the attack.
Here is a more concise version in paragraph form, preserving the citations and covering both voice and video modalities:

Real-time deepfakes (RT-DFs) have rapidly evolved from a theoretical risk into a practical threat, with documented harms across financial fraud, political deception, corporate espionage, and extortion. In finance, attackers have used real-time voice cloning to impersonate executives and induce unauthorized transfers, including an early case in which scammers mimicked a CEO’s voice to steal \$243,000 \cite{Fraudste87:online}, as well as a 2024 incident in which a Hong Kong employee was deceived during a deepfake video call featuring fabricated likenesses of the company’s CFO and other staff, leading to a loss of roughly \$25.6 million \cite{Millions20:online}. RT-DFs have also been used in large-scale theft, such as a \$35 million bank heist carried out through fraudulent audio calls impersonating a company director \cite{Fraudste98:online}.

Beyond direct financial theft, RT-DFs enable manipulation, infiltration, and intelligence gathering. Politically, they can be used to impersonate influential figures and interfere with diplomatic or public processes \cite{Sophisti0:online}; this vulnerability was exposed when senior European MPs joined Zoom meetings with individuals posing as Russian opposition figures \cite{European87:online}. In corporate settings, the FBI has warned that cybercriminals are using deepfakes in job interviews to obtain remote positions and gain access to sensitive systems and information \cite{Internet56:online}. RT-DFs have also been deployed for executive targeting: in May 2024, attackers reportedly used a deepfake of WPP CEO Mark Read’s voice and likeness in a Microsoft Teams meeting to solicit money and personal information from other executives \cite{Binancee50:online}.

RT-DFs are also expanding into coercive scams designed to provoke panic and extract payment. One notable example involved an Arizona mother who received a call in which scammers used AI-cloned audio of her daughter’s voice to simulate a kidnapping and demand a \$1 million ransom \cite{USmother64:online}. Taken together, these cases show that RT-DF threats now span both voice and video modalities and are becoming more diverse, scalable, and operationally sophisticated. Their progression from isolated fraud attempts to coordinated, high-impact attacks highlights the urgent need for stronger detection, authentication, and prevention mechanisms.

\subsection{The Gap in Current Defenses}

While numerous methods have been proposed for detecting deepfakes \cite{mirsky2021creation,almutairi2022review}, the rapidly evolving nature of this technology presents significant challenges to existing defense mechanisms. Current approaches typically rely on deep learning models to either (1) identify artifacts or inconsistencies in generated media, or (2) detect forensic evidence such as latent noise patterns. However, these methods face two fundamental problems that limit their long-term effectiveness:
\begin{description}[leftmargin=.5cm]
    \item[Longevity.]  Methods that identify semantic errors or artifacts operate under the assumption that deepfake quality will remain relatively stable. However, evidence clearly demonstrates that deepfake quality is rapidly improving \cite{masood2023deepfakes}. Consequently, artifact-based methods face a high risk of obsolescence within a short timeframe, as deepfake technology outpaces their detection capabilities.
    \item[Evasion.]  Methods relying on latent noise patterns are vulnerable to evasion through simple post-processing techniques. For instance, a deepfake can be passed through a low-pass filter, undergo compression, or be subjected to additive noise. Critically, these processes are common in audio and video calls, potentially eliminating forensic evidence without additional effort from the attacker.
\end{description}

These weaknesses are amplified in the real-time setting. The signal degradation that pervades a live call already mimics the post-processing an evader would otherwise have to apply, while the strict latency budget of an interactive conversation rules out the heavyweight forensic analysis that some defenses depend on. Detecting RT-DFs therefore demands a fundamentally different posture, one that does not wait for the attacker's content to betray itself.

\section{Real-Time DF-CAPTCHA}
\begin{figure}[t]
   \centering
   \includegraphics[width=0.7\columnwidth]{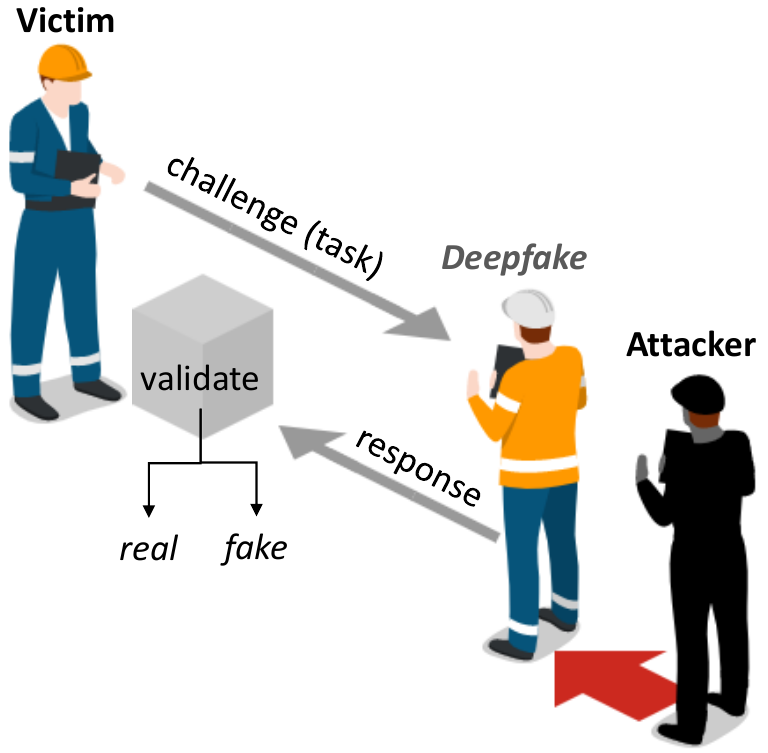}
   \caption{Overview of the proposed defense against fake calls: the victim requests the caller to perform a task that is difficult for deepfake models to execute accurately. If the response is distorted or fails to address the task, the caller is likely a deepfake.}
   \label{fig:overview}
\end{figure}

Rather than wait for an artifact to appear, we flip the dynamic and force the adversary into the spotlight. We introduce Deepfake-CAPTCHA (\textbf{DF-CAPTCHA}): a system that automatically detects deepfake calls through challenge-response analysis (illustrated in Fig.~\ref{fig:overview}). DF-CAPTCHA actively engages the caller by issuing a specific task (the \textit{challenge}), deliberately chosen to be effortless for a human yet outside the operating envelope of a real-time deepfake pipeline. Our key insight is that, unlike general-purpose generative models, RT-DF pipelines are tightly constrained: every frame must preserve the target's identity, geometry, and temporal coherence under a strict latency budget. This rigidity makes the pipeline brittle along many axes and exposes a rich attack surface for the defender to exploit. When a deepfake attempts the challenge, the resulting content (the \textit{response}) collapses in ways that are conspicuous to an anomaly detector, a downstream classifier, or even the intended victim.

In our previous work~\cite{yasur2023deepfake}, we focused exclusively on RT-DF attacks involving audio-based voice cloning. In this paper, we extend that framework to video-based attacks, enabling a comprehensive analysis of RT-DF threats \textit{across both audio and video modalities}. Concretely, our extension contributes (1) a threat analysis of current RT-DF face-impersonation technologies supported by a user study, (2) the proposal and in-depth evaluation of novel defensive challenges for real-time face impersonation, and (3) insights into the structural limitations of existing RT-DF face-impersonation pipelines. Together, these allow us to assess DF-CAPTCHA across a substantially broader range of real-time deepfake scenarios.

Our results show that DF-CAPTCHA substantially improves the performance of state-of-the-art deepfake detectors in both modalities, offering a robust and adaptive defense against increasingly sophisticated RT-DF attacks. Beyond improving detection accuracy, it provides a scalable framework that can evolve alongside the underlying deepfake technology.

\subsection{Contributions}
In summary, this work makes the following contributions:
\begin{itemize}
    \item \textbf{The first active defense against RT-DFs.} Our approach (i) provides stronger detection guarantees than passive, artifact-based methods, and (ii) offers improved longevity through an extensible challenge set that can adapt as deepfake technology evolves.

    \item \textbf{A formal definition and analysis of DF-CAPTCHA.} We define what constitutes a strong deepfake CAPTCHA, formalize the four constraints a challenge must impose on the adversary, and present an initial catalogue of CAPTCHAs together with a security and usability analysis.

    \item \textbf{The first comprehensive threat evaluation of RT-DF technology with real users.} We evaluate four RT-DF face-replacement models with 38 volunteers and five  voice-cloning models with 41 volunteers, providing the first user-study-grounded view of the current RT-DF threat landscape.

    \item \textbf{An end-to-end performance and robustness evaluation.} We provide an in-depth analysis of how DF-CAPTCHA performs in both the video and audio domains, and of its resilience against an evasive adversary.
\end{itemize}
These contributions collectively advance deepfake detection by offering a novel, adaptable, and thoroughly evaluated approach to the evolving threat of real-time deepfakes.

\section{Literature review}
\subsection{Audio Deepfake Detection}

Audio deepfake detection (ADD) systems typically follow a three-step pipeline:
\begin{enumerate}
    \item Convert audio clip $a$ into frames $a^{(1)},... a^{(n)}$
    \item Extract feature representations $x^{(1)},... x^{(n)}$ (e.g., STFT, MFCC, CQCC \cite{lei2020siamese, lai2019assert}, or raw waveform)
    \item Pass frames through a detector to predict authenticity
\end{enumerate}
ADD systems use either classifiers \cite{kawa2022specrnet, rawnet2, khochare2022deep} or anomaly detectors \cite{khalid2020oc}. Classifiers are trained on labeled real and fake audio data, automatically identifying distinguishing features \cite{agarwal2020detecting}. However, they assume all deepfake types are in the training set, requiring retraining for new technologies. Models range from classical machine learning \cite{borrelli2021synthetic} to deep learning architectures \cite{zhang2021one, camacho2021fake, liu2021identification, arif2021voice}.
Anomaly detectors, trained only on real voice data, flag abnormal patterns. They use voice recognition embeddings \cite{pianese2022deepfake} or one-class models like OC-SVMs and GMMs \cite{todisco2019asvspoof, zhang2021one, khalid2020oc}.
Despite efforts to improve generalization \cite{kawa2022attack}, ADD systems still struggle with new audio distributions and novel deepfake technologies \cite{muller2022does}.
A comprehensive review of modern ADD systems can be found in \cite{almutairi2022review}.

\begin{description}[leftmargin=.0cm]
\item[Fraudulent Call Prevention Services.] Existing fraudulent call prevention services use blacklists and statistical information surrounding a caller's phone number to identify and block suspicious calls (e.g., Truecaller, Robo Killer and Nomorobo). These approaches are only effective against general phishing attacks where many fraudulent calls are made from the same numbers or same telephony region. However, in our threat model, we focus on spear phishing attacks where specific victims are targeted and the attacks are crafted for those victims. In a spear phishing attack the caller only needs to make one call from a 'clean' phone number once to achieve his or her goal, making existing prevention methods less effective. In contrast, our DF-CAPTCHA method (1) examines the content of the call (i.e., analyzes the caller's voice) and (2) does not require any prior knowledge of potential callers (voice identities, phone numbers, etc.) making it a suitable defense against spear phishing attacks which use RT-DFs.
\end{description}

\subsection{Video Deepfake Detection}
Video deepfake detection (VDD) systems typically follow a four-step pipeline:
\begin{enumerate}
    \item Convert video clip $a$ into frames $a^{(1)},... a^{(n)}$
    \item Detect and locate human faces in each frame
    \item Extract features manually or automatically
    \item Pass frames through a detector to predict authenticity
\end{enumerate}

 Features extracted include facial landmarks \cite{li2021deepfake, yang2019exposing}, movement patterns using optical flow \cite{amerini2019deepfake}, or deep learning embeddings \cite{afchar2018mesonet, wodajo2021deepfake}.

VDD approaches are broadly categorized into handcrafted features and deep learning techniques. Handcrafted methods include 3D head position estimation \cite{yang2019exposing}, multimedia stream descriptors \cite{guera2019we}, biological signals like heart rate \cite{ciftci2020fakecatcher}, and eye-blinking analysis \cite{jung2020deepvision}. Deep learning approaches encompass CNN models for facial landmark analysis \cite{li2018exposing}, CNN-RNN architectures for frame-level detection \cite{guera2018deepfake, li2018ictu}, multi-task CNNs with RNNs \cite{montserrat2020deepfakes}, and temporal information learning \cite{de2020deepfake}. Advanced techniques combine facial and behavioral biometrics \cite{agarwal2020detecting}, employ heart rate measurement with Neural-ODEs \cite{fernandes2019predicting}, and utilize multi-scale self-texture attention \cite{yang2021msta}.

These methods aim to distinguish between real and fake videos by analyzing various aspects of video content. However, challenges persist in generalizing to new types of deepfakes and handling different video qualities. For a comprehensive review of modern VDD systems, readers can refer to \cite{masood2023deepfakes}.

\subsection{Limitations of Current Defenses}
Current deepfake detection methods in both video and audio domains primarily employ passive analysis, examining content without engaging the caller. In contrast, our proposed DF-CAPTCHA method introduces an active defense paradigm. By compelling the deepfake function to generate content beyond its capabilities, DF-CAPTCHA induces more pronounced artifacts, enhancing detection efficacy.
DF-CAPTCHA's active approach offers significant advantages over existing methods. It improves detection rates and ensures longevity against evolving attacks by targeting specific limitations of deepfake technologies. Moreover, the challenge-response nature of DF-CAPTCHA enables precise localization of artifacts within media streams. This targeted analysis increases efficiency by focusing on specific segments rather than entire streams, overcoming limitations associated with fixed-size segment analysis methods \cite{zhang2021one, rawnet2, guera2018deepfake, muller2022does}.
By actively probing deepfake limitations and leveraging precise artifact localization, DF-CAPTCHA presents a more robust, efficient, and adaptable solution to the challenges posed by real-time deepfakes in both audio and video domains.

\label{sec:related_work}
\section{Methodology}
This section examines the limitations of RT-DFs and uses them to define how D-CAPTCHAs work.

\subsection{RT-DF Limitations}\label{subsec:limitations}
Current RT-DF models generate only content within the scope of the task on which they were trained. For example, a model trained to reenact $t$'s face in a roughly frontal pose or generate $t$'s voice in a calm speaking style will generally fail on other content. This is evident in facial reenactment systems such as \cite{NIPS2019_8935} and \cite{perov2020deepfacelab}: they perform well on frontal faces but cannot generate the back of the target's head. Likewise, audio RT-DFs struggle to identify and reproduce sounds outside the regime emphasized by the training data, loss functions, and overall pipeline, such as when the system is optimized for normal speech.

An \textbf{ideal RT-DF} would generate realistic, identity-faithful content of $t$ performing an \textit{arbitrary} task. Existing RT-DFs are not ideal because they are trained for narrower tasks, which allows $f_t$ to better preserve identity and realism in $x_g$ when driven by $x_s$. Thus, even if \textit{out-of-domain} tasks can be anticipated, $f_t$ cannot realistically be trained to reproduce them all. This limitation arises from both technology and practicality.

\subsubsection{Technology}
These limitations stem from the fact that \textit{current} technology cannot yet realize the \textit{ideal} RT-DF.

\begin{description}[leftmargin=.5cm]
    \item[Inference Speed.]
    The rate of audio-frame generation depends on the efficiency of the deepfake pipeline and the model architecture. Supporting a broad range of tasks would require many more parameters\footnote{For reference, the voice deepfake model StarGAN \cite{StarGANv2-VC}, a audio-based RT-DF, has about 53 million parameters. In contrast, models that generate arbitrary content, such as DALL-E 2 and Imagen, use 3.5--4.6 billion parameters. Moreover, methods such as stable diffusion require multiple passes.} and likely more complex feature extractors. For example, current RT-DFs would need higher-resolution STFTs and MFCCs to cover a wider frequency range.

    \item[Feature Representation.]
    To learn and reproduce patterns in \( x_s \), the model must extract meaningful representations from the media. In voice, salient patterns often lie in lower frequencies and have a relatively stable spectral envelope compared with sounds such as singing or clapping. Current pipelines often rely on compressed features such as MFCCs or STFTs sampled at relatively low rates (e.g., 16--24 kHz \cite{almutairi2022review}). Capturing a wider frequency range would require higher-resolution input, but this greatly increases model complexity and training difficulty. In face generation, preprocessing is also critical before the model sees the input. This often includes detecting, cropping, and aligning the driver's face. Some pipelines reduce preprocessing to facial landmarks only, which may omit subtle cues such as muscle contractions and thereby limit nuanced generation.

    \item[Training.]
    Training requires a loss function to guide optimization. Modern RT-DF systems typically use at least two: one for realism (e.g., adversarial loss) and one for preserving the identity of $t$ in $x_g$ (e.g., perceptual loss) \cite{mirsky2021creation}. Supporting additional tasks would likely require additional losses, but these objectives compete during optimization, so some aspects degrade. More losses can also make convergence harder. In addition, $x_s$ may contain multiple identities or irrelevant content, such as multiple faces and physical interactions in video, or multiple voices and background sounds such as music in audio. To use such a source signal, the model would need to isolate the relevant components and then recombine them correctly in $x_g$; otherwise, the output will be corrupted or the missing context will be noticeable to the victim. To the best of our knowledge, this remains an open problem.
\end{description}

\subsubsection{Resources}
These limitations arise when the desired output may be achievable with current technology, but obtaining it is prohibitively expensive or impractical.

\begin{description}[leftmargin=.5cm]
    \item[Data Collection.]
    Producing a high-quality RT-DF of $t$ requires substantial audio data from $t$ (e.g., \cite{StarGANv2-VC} requires 20--30 minutes). However, it is impractical for an attacker to collect audio of $t$ performing many specific tasks beyond ordinary speech. Lower quality may be possible with zero-shot learning, but this still requires (1) a broad dataset covering many possible tasks and (2) a model that generalizes those samples to new identities.

    \item[Knowledge.]
    Building a system that can handle even a subset of arbitrary tasks requires significant expertise in generative deep learning. This raises the barrier for casual attackers, though not necessarily for advanced adversaries.

    \item[Labeling.]
    Annotating and labeling large datasets is expensive and time-consuming, and the burden increases with the number of classes (tasks).

    \item[Assets.]
    An ideal RT-DF capable of arbitrary tasks would likely be highly complex. Running such a model in real time would require a powerful GPU, which may be prohibitively expensive or may not yet exist.
\end{description}

\subsubsection{Outlook on RT-DF Limitations}
The limitations described here apply to current RT-DF systems. Although they are difficult to overcome, future RT-DF technologies may not share all of them. Still, we expect that some constraints, especially data collection and training, will remain relevant for novel systems in the near future.

Therefore, to gain an advantage over the adversary, defenses should exploit RT-DF limitations whenever possible.

\subsection{Defining DF-CAPTCHA}

The concept of CAPTCHA, as defined by Ahn et al. \cite{ahn2003captcha}, is ``\textit{a cryptographic protocol whose underlying hardness assumption is based on an AI problem.}'' This protocol traditionally follows a challenge-response procedure between a server $A$ (the verifier) and a client $B$ (the prover), structured as follows:

\begin{enumerate}
    \item $A\rightarrow B: c$ (Server sends challenge $c$ to client)
    \item $B\rightarrow A: r_c$ (Client sends response $r_c$ to challenge $c$)
    \item $A: V(r_c) \in \{pass, fail\}$ (Server verifies if $r_c$ resolves challenge $c$)
\end{enumerate}

Conventional CAPTCHAs, such as reCAPTCHA, aim to differentiate humans from bots by presenting tasks that are simple for humans but challenging for automated systems (e.g., decoding distorted text).

In contrast, our proposed Deepfake-CAPTCHA (DF-CAPTCHA) introduces a paradigm shift by challenging the client to \textbf{create} content under specific constraints:

\begin{enumerate}
    \item \textbf{Realism}: The generated content must appear authentic to both human perception and machine learning models.
    \item \textbf{Identity}: The content must accurately reflect a given identity $t$.
    \item \textbf{Task Complexity}: The content must depict identity $t$ performing an arbitrary task that is difficult for current deepfake technologies to generate convincingly.
    \item \textbf{Real-Time Generation}: The content must be produced in real-time, without significant delay.
\end{enumerate}

This novel approach leverages the current limitations of Real-Time Deepfake (RT-DF) technologies. While existing RT-DF systems struggle to satisfy these constraints simultaneously, humans can easily meet these requirements. The 'hardness' of a DF-CAPTCHA is thus directly correlated with the current limitations of RT-DF technology.

Importantly, like modern CAPTCHA systems, DF-CAPTCHA is designed with extensibility in mind. As RT-DF technologies evolve, new challenges can be seamlessly integrated into the system, ensuring its effectiveness against future threats. This adaptability provides our system with a crucial advantage in the ongoing arms race between deepfake creation and detection technologies.

By framing deepfake detection as a challenge-response protocol, DF-CAPTCHA not only offers a robust method for identifying synthetic content but also establishes a flexible framework that can evolve alongside advancements in deepfake technology. This approach represents a significant step forward in active defense strategies against real-time deepfakes in both audio and video domains.

\subsubsection{Creating a Challenge}
A challenge in the DF-CAPTCHA system is designed to assess a caller's ability to create content that simultaneously satisfies four key constraints: realism, identity, task complexity, and time. Each of these constraints serves a specific purpose in differentiating between genuine human responses and deepfake attempts:

\begin{description}
    \item[Realism] constraints are crucial for ensuring the absence of latent or semantic anomalies in the response. This helps in identifying synthetic content that may appear superficially convincing but contains subtle inconsistencies.
    \item[Identity] constraints are designed to push deepfake models beyond their current capabilities, exploiting known limitations in their ability to generate certain types of content.
 
    \item[Task]  constraints are designed to push deepfake models beyond their current capabilities, exploiting known limitations in their ability to generate certain types of content.
    \item[Time] constraint ensure the use of real-time deepfake (RT-DF) models, preventing attackers from switching to more sophisticated offline models.
\end{description}

The core of each challenge is a specific task that the caller must perform. We denote a specific task as $T$, where, for example, $T=hum$ might represent the task "hum a specific song." The set of all possible challenges for a given task $T$ is denoted as $C_T$. For instance, $C_{hum}$ would encompass all possible requests for different songs to be hummed.

The challenge selection process follows a two-step randomization procedure:
Two random seeds, $z_0$ and $z_1$, are generated.
$z_0$ is used to select a random task $T$ from the set of available tasks.
$z_1$ is then used to select a random challenge $c$ from the set $C_T$.

This randomized selection process ensures unpredictability in the challenges, making it difficult for attackers to anticipate and prepare for specific tasks. By combining these carefully designed constraints with a randomized selection mechanism, the DF-CAPTCHA system creates challenges that are easily surmountable by genuine human callers but present significant obstacles for current RT-DF technologies.

Table \ref{tab:captchas_combined} presents a selection of tasks that can be employed in DF-CAPTCHA challenges. Our evaluation assumes that the Real-Time Deepfake (RT-DF) under test has been optimized for maximum performance on a single task: regular talking in audio and face video. This assumption allows us to assess the effectiveness of diverse challenges against a highly specialized RT-DF system.

We conducted an assessment of these tasks based on observations from various RT-DF models. Our evaluation considers key factors:

\begin{description}
    \item[Effectiveness:]  This metric quantifies the difficulty for a modern RT-DF to successfully create a convincing deepfake of target identity $t$ given the specific task constraints. Higher hardness indicates a more challenging task for the RT-DF system.
    \item[Robustness:] This factor identifies potential evasion strategies that an adversary might employ for each task. We consider two primary evasion tactics: \textit{Bypass:} The attacker disables the RT-DF and directly interacts with our system.
    \textit{Mix:} The attacker combines the RT-DF output ($a_g$) with additional audio sources. For example, to evade a 'talk \& clap' challenge, an attacker might create $a'g = a_g + a_\text{clap}$, where $a_\text{clap}$ is captured from a separate microphone to avoid disrupting the RT-DF process (i.e., $f_t(a_s + a_m)$).
 
    \item[Sophistication:] We evaluate the efficacy of each challenge against two levels of attackers: 
\textit{Naive Attacker:} Uses existing datasets and limited samples of $t$ to train $f_t$, and routes all audio through $f_t$ (e.g., using an unmodified library from GitHub).
\textit{Advanced Attacker:} Collects a practical amount of samples on $t$ (e.g., 20 minutes) and can mix additional sources into $a_g$.

\end{description}

Our analysis reveals that a robust challenge is characterized by a random task $T$ and a random challenge $c$ drawn from $C_T$, which collectively present significant difficulties for the adversary across all four constraints (realism, identity, task complexity, and time).
This comprehensive evaluation framework enables us to systematically assess and refine the challenges used in the DF-CAPTCHA system, ensuring their effectiveness against a range of potential attack strategies. By considering both the limitations of current RT-DF technologies and the capabilities of different attacker profiles, we can design challenges that maintain a high level of security while remaining feasible for genuine human responses.

\subsubsection{Verifying a Challenge}\label{subsubsec:verify}

To determine whether $V(r_c)=pass$ or $fail$, we must verify if $r_c$ adheres to the \textbf{realism}, \textbf{identity}, \textbf{task}, and \textbf{time} constraints. While all four constraints can be verified by a human (either a moderator or the victim), human verification may be unreliable due to lack of confidence or susceptibility to social engineering. Therefore, we propose an automated verification method for each constraint that doesn't require prior knowledge of the target identity $t$.

The verification process for $r_c$ involves validating each constraint separately:

\begin{description}[leftmargin=.5cm]
    \item[Realism Verification ($\mathcal{R}$).] This check identifies distortions and artifacts in $r_c$ that may occur when an RT-DF operates outside its capabilities or when a low-quality RT-DF is used. We employ existing anomaly detectors and deepfake classifiers to assess the content's realism. $\mathcal{R}$ outputs a score in the range $[0,\infty)$ or $[0,1]$, indicating the degree of unrealism in $r_c$.
    
    \item[Identity Verification ($\mathcal{I}$).] To verify that $r_c$ maintains the claimed identity $t$, we:
    \begin{enumerate}
        \item Collect a short sample $a_t$ of the caller before the challenge, with the victim confirming the identity.
        \item Use a zero-shot recognition model to compare the identities in $a_t$ and $r_c$.
    \end{enumerate}
    This two-step process prevents identity switching after the challenge. $\mathcal{I}$ outputs a similarity score between $a_t$ and $r_c$. An alternative approach using continuous verification could be employed, although it would be more resource-intensive.
    
    \item[Task Verification ($\mathcal{C}$).] This check ensures that $r_c$ contains the requested task, addressing scenarios where:
    \begin{enumerate}
        \item The model fails to generate the required content.
        \item The attacker attempts to evade detection by performing a different task or no task at all.
    \end{enumerate}
    We employ a machine learning classifier for this verification. $\mathcal{C}$ outputs the probability that $r_c$ \textit{does not} contain the specified task.
 
    \item[Time Verification ($\mathcal{T}$).] We verify the time constraint by ensuring that the first frame of $r_c$ is received within approximately 1 second after the challenge instructions are given. $\mathcal{T}$ outputs the measured time delay, denoted as $d$.
\end{description}

The overall validation of $r_c$ is determined by comparing the outputs of all four algorithms ($\mathcal{T},\mathcal{R},\mathcal{I},\mathcal{C}$) against their respective thresholds ($\phi_1,\phi_2,\phi_3,\phi_4$), each tuned appropriately. We consider $r_c$ valid if none of the algorithm outputs exceed their thresholds, and invalid if any threshold is exceeded.
In summary, validation is performed as follows:
\begin{equation}
    V(r_c) = 
    \begin{cases}
      pass, \hspace{1em} \mathcal{T}(d)< \phi_1, \mathcal{R}(r_c)< \phi_2, \\
      \hspace{3em} \mathcal{I}(r_c,a_t)< \phi_3, \mathcal{C}(r_c,c) < \phi_4    \\
      fail, \hspace{1em} \text{else}
   %     Valid \&\text{a(u_i,c_j)>TreshC \land i(u_i,c_j)>TreshI}\\
 %       Unvalid \&\text{$else$}\
    \end{cases}
\end{equation}
While the false reject rate can be adjusted by weighting the contribution of each constraint, it's important to note that such adjustments may compromise the system's security. Therefore, careful consideration must be given to balancing user experience with robust security measures.

This comprehensive verification process ensures that the DF-CAPTCHA system can effectively distinguish between genuine human responses and sophisticated deepfake attempts across multiple dimensions of authenticity.

\begin{figure*}[t]
    \centering
    \includegraphics[width=.9\textwidth]{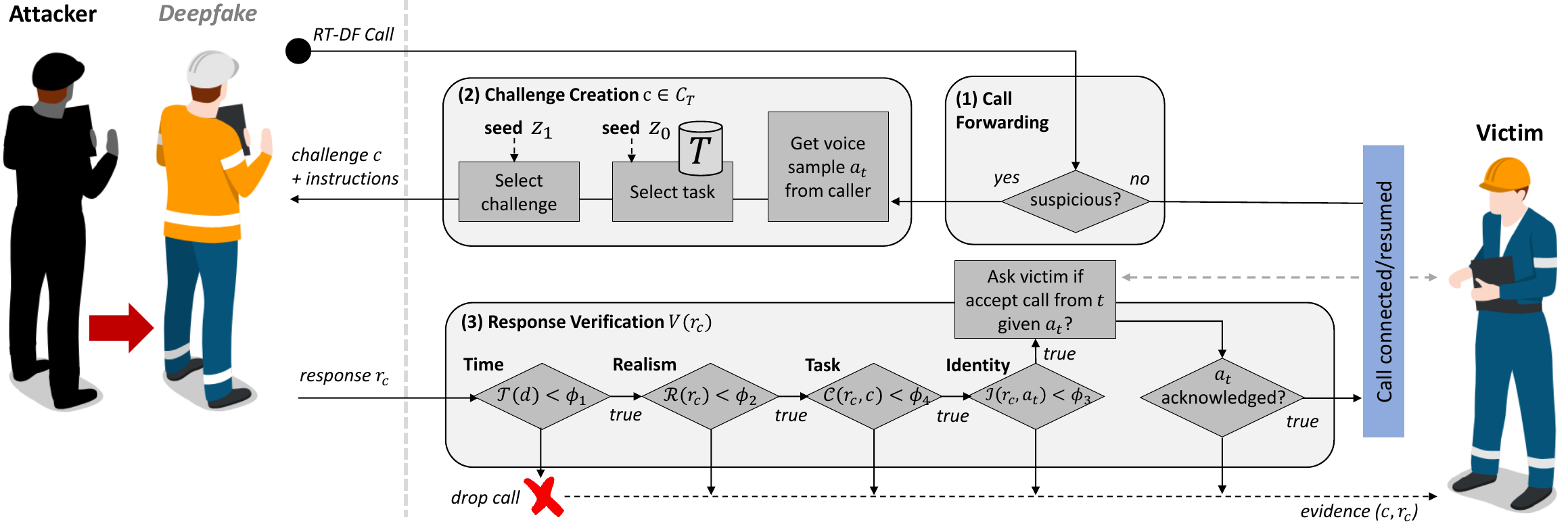}
    \caption{An overview of the proposed DF-CAPTCHA system: (1) Calls are forwarded to the system using a blacklist, whitelist, policy or the victim's intuition, (2) a random DF-CAPTCHA $c$ with accompanying instructions is generated and send to the caller as a challenge, (3) the response $r_c$ is verified against the four constraints (time, realism, identity, task) and if all four pass then the call is connected/resumed. Otherwise, the call is dropped and evidence is provided to the victim.}
    \label{fig:framework}
\end{figure*}

\input{captcha_table_all.tex}

\subsection{Detection Framework}

DF-CAPTCHA protects users from fake callers through three stages: call forwarding, challenge creation, and response verification.

\noindent\textbf{1: Call Forwarding. }
Calls can be routed to DF-CAPTCHA in several ways. In high-risk settings, all calls may be screened. Alternatively, screening can be triggered by blacklists, policies for callers outside the user’s address book, unexpected or unknown calls, suspicious conversations, unusual call behavior, or before sensitive discussions.

\noindent\textbf{2: Challenge Creation. }
A random challenge $c$ is sampled from the available challenge set, and instructions are generated describing the required action and its start cue. These instructions are delivered to the caller, e.g., via Text-to-Speech (TTS).

At the start of the challenge, the caller states their name. This sample, denoted $a_t$, is shared with the victim for acknowledgment and used for identity verification.\footnote{This prevents an attacker from simply disabling the RT-DF during the challenge and responding with their real voice or face.} The instructions are then given, followed by a tone. The delay between the tone and the first response is measured and included in $r_c$ for time verification ($\mathcal{T}$). After a fixed interval, the response is recorded as $r_c$ and passed to the verification stage.

\noindent\textbf{3: Response Verification. }
The response $r_c$ and its timing data are evaluated by $\mathcal{T}, \mathcal{R}, \mathcal{C},$ and $\mathcal{I}$. If all outputs are below their respective thresholds, $a_t$ is played to the user. If the user accepts the claimed identity, the DF-CAPTCHA passes and the call is connected or resumed.

If any component exceeds its threshold, the call is dropped and evidence is presented to the user. This may include which constraint(s) failed, by how much, and playback of $a_t$, $c$, and $r_c$ when appropriate.

Although the execution order is not semantically important, we suggest $\mathcal{T}\rightarrow\mathcal{R}\rightarrow\mathcal{C}\rightarrow\mathcal{I}$ to reduce unnecessary computation. In higher-security settings, multiple DF-CAPTCHAs can be issued and verified to reduce the false negative rate.

\subsection{Deployment}
The framework can be deployed either on-device, such as as a phone app, or in the cloud. This supports personal call screening as well as organizational settings, including call centers and online meeting rooms, where callers can be screened before being admitted to sensitive conversations such as confidential Zoom meetings \cite{European87:online,Binancee98:online}.

\noindent\textbf{Usability \& Limitations. }
DF-CAPTCHA provides a practical active defense against real-time impersonation, but it also has important limitations. It is designed specifically for Real-Time Deepfakes (RT-DFs), making it effective against live impersonation attempts but not against pre-recorded deepfake content. As an active defense, it must also balance security with convenience: poorly calibrated challenges may become intrusive or time-consuming for legitimate callers. Nevertheless, this trade-off is often acceptable in high-security settings where call authenticity is critical, such as confidential business or sensitive personal communications.

The system can operate either autonomously or manually, depending on the user’s needs and security requirements, which improves usability across a range of scenarios. A video demonstration of the system is available online.\footnote{\url{https://youtu.be/izSeINj53hg}} DF-CAPTCHA also relies on deep learning models for realism ($\mathcal{R}$), identity ($\mathcal{I}$), and task ($\mathcal{C}$) verification. Although such models are, in principle, vulnerable to adversarial attacks \cite{carlini2020evading}, several factors improve robustness. Calls typically pass through noisy and compressed channels, which can weaken adversarial perturbations; generating adversarial examples in real time substantially increases attack complexity and computational cost; and the verification models are likely black boxes from the attacker’s perspective, making them difficult to query or manipulate.

Overall, while DF-CAPTCHA is limited in scope and may introduce some user friction, it remains a strong mechanism for protecting live call authenticity against increasingly sophisticated digital impersonation. Its flexibility and robustness make it especially suitable for high-security environments where the cost of a false acceptance outweighs the inconvenience of additional verification.

\label{sec:methodology}
\section{Threat Analysis}
In this chapter, we evaluate the threat posed by Real-Time Deepfakes (RT-DFs) by assessing the quality and capabilities of various RT-DF models through user studies. Our experimental study was designed to gauge both the performance of current RT-DF algorithms and the ability of humans to identify manipulated content in realistic settings. Participants were exposed to authentic and manipulated media in audio and video scenarios and were asked to judge realism, identity, trust, and authenticity. This allows us to characterize the present threat level of RT-DFs while also measuring how well human observers can detect them. All of our experiments were approved by our university’s ethics committee.

Because this paper is an extension of our original audio-focused work, the audio results reported here summarize and build on our original findings, while the video results constitute the new experimental component introduced in this journal version. To present a unified view of the RT-DF threat landscape, we describe the audio and video evaluations in parallel.

\subsection{Audio and Video RT-DFs}
We evaluate RT-DFs in both the audio and video domains using separate user studies. In the audio domain, we assess the quality of five RT-DF voice cloning models from the perspective of 41 volunteers. In the video domain, we assess the quality of four different RT-DF face-manipulation models based on feedback from 38 volunteers. In both studies, the participants were university students with an average age of 25, and most were not studying computer science or related disciplines. In the audio study, 18 participants were women and 23 were men. In the video study, the group comprised 15 women and 23 men. To avoid bias in the video experiments, the volunteers were not informed that the study concerned deepfakes or cybersecurity until the end of the experiment.

\subsection{Experiment Setup}
\subsubsection{RT-DF Models}
For the audio evaluation, we surveyed 25 voice cloning papers published over the last four years that can process audio in real time as a sequence of frames. Of these, we selected the four recent works that published source code: \texttt{AdaIN-VC} \cite{AdaIN}, \texttt{MediumVC} \cite{MediumVC}, \texttt{FragmentVC} \cite{FragmentVC}, and \texttt{StarGANv2-VC} \cite{StarGANv2-VC}. We also included \texttt{ASSEM-VC} \cite{Assem-VC} as an additional comparison. All audio clips in this experiment were generated using the pre-trained models provided by the original authors. To simulate a realistic telephony setting, the clips were passed through a phone filter, i.e., a band-pass filter in the 0.3--3 KHz voice range \cite{phone_filter}.

For the video evaluation, we reviewed academic papers and popular GitHub community projects to identify the latest technologies that (1) operate in real time and (2) provide source code to ensure reproducibility. We selected four models: \texttt{DeepFaceLab} \cite{perov2020deepfacelab}, \texttt{FaceFusion} \cite{facefusion}, \texttt{SimSwap} \cite{DBLP:conf/mm/ChenCNG20}, and Roop \cite{roop}. For our experiments, all video clips were generated using the pre-trained models provided by the original authors.

\subsubsection{Experiments}
To quantify the threat of RT-DFs, we conducted two experiments in audio and three in video.

\paragraph{Audio experiments.}
Following our original audio evaluation, we performed two experiments on a group of 41 volunteers:
\begin{description}[leftmargin=.3cm]
    \item[EXP1a - Quality.] The goal of the first experiment was to determine how easy it is to identify an RT-DF in the best-case scenario, namely when the victim is already expecting a deepfake.
    \item[EXP1b - Identity.] The goal of the second experiment was to determine how well RT-DF models can clone identities.
\end{description}

In \textbf{EXP1a}, volunteers were asked to rate the quality of the voices on a scale of 1--5 (1: fake, 5: real). There were 90 audio clips presented in random order: 30 real and 60 fake, with 12 clips from each of the five RT-DF models. The clips were approximately 4--7 seconds long each.

In \textbf{EXP1b}, we selected the top two models from EXP1a. For each model, we repeated the following trial eight times. First, the volunteer listened to two real samples of the target identity as a baseline. Then we played two real and two fake samples in random order and asked the volunteer to rate how similar the speakers sounded compared to the speaker in the baseline.

If a model has a positive mean opinion score (MOS) in both EXP1a and EXP1b, then it poses a considerable threat, since it can (1) synthesize high-quality speech, (2) that sounds like the target, and (3) do so in real time. The distribution of ratings across the five audio RT-DF models and real recordings is shown in Fig.~\ref{fig:exp1.1}.

\paragraph{Video experiments.}
To quantify the threat of RT-DFs in video, we performed three experiments on a group of 38 volunteers:
\begin{description}[leftmargin=.3cm]
    \item[EXP2a - Authenticity.] The goal of the first experiment was to assess whether a victim would perceive a deepfake video as authentic \textit{without} anticipating a deepfake. In other words, \textit{how likely would someone fall for this attack?}
    \item[EXP2b - Detection.] This experiment aimed to assess participants’ ability to discriminate between deepfake-generated videos and authentic footage while anticipating that some of the videos would be fake. In other words, does an informed decision affect one’s susceptibility?
    \item[EXP2c - Quality.] This phase of the experiment aimed to evaluate the perceptual quality and believability of synthetically generated videos. In other words, which technology is the best?
\end{description}

In \textbf{EXP2a}, participants were shown a video of a student introducing themselves using the standard script: ``\textit{Hello, my name is [Name], and I study at [University Name]}.'' The participants, who were unfamiliar with the identities in the videos, were asked to rate their trust in the person on a scale of 1--5. The participant pool was divided into two groups. For each identity, one group viewed the authentic recording and the other viewed a deepfake version of the same identity. At this stage, participants were unaware that some videos had been generated by deepfake algorithms or that the study was evaluating deepfake perception. The goal was to determine whether there is a significant difference in how people perceive and trust authentic versus deepfake videos.

In \textbf{EXP2b}, participants were informed that the videos they were about to view contained both authentic and synthetic content. The synthetic videos were generated using \texttt{DeepFaceLab}, the algorithm that produced the most convincing results in our earlier evaluation. Participants were then shown a mix of authentic and deepfake videos and were asked to rate authenticity on a 5-point Likert scale: (1) Certain the video is authentic, (2) Believe the video is authentic, (3) Uncertain about the video’s authenticity, (4) Believe the video is synthetic, and (5) Certain the video is synthetic. The objective was to test whether people could accurately distinguish between authentic and deepfake videos when explicitly primed to do so.

In \textbf{EXP2c}, we compared the previously selected deepfake algorithms directly. Participants were shown five videos generated by each deepfake algorithm and five authentic recordings. For each video, they rated two attributes. The first was perceived authenticity on a 1--5 scale, where 1 indicated ``easily identifiable as AI-generated'' and 5 indicated ``indistinguishable from a genuine recording.'' The second was overall video quality, focusing on visual clarity and fidelity, again on a 1--5 scale, where 1 represented ``low quality with noticeable distortions'' and 5 represented ``high quality with no apparent flaws.'' This design allowed us to distinguish between an algorithm’s ability to generate convincing fakes and its ability to produce high-quality visual output. The distributions of ratings across the evaluated video RT-DF models and real videos are shown in Fig.~\ref{fig:exp3.1}.

\subsection{Experiment Results}
\paragraph{Audio results.}
The audio results reported here summarize our original findings and serve as the historical baseline for the multimodal threat analysis in this extension paper.

\textbf{EXP1a.} To analyze the quality (realism) of the models, we compared the MOS scores of the deepfake audio to the MOS of the real audio, where both were scored blindly. In Fig.~\ref{fig:exp1.1}, we plot the distribution of each model’s MOS compared to real audio. Roughly 20--50\% of the volunteers gave the RT-DF audio a positive score, with \texttt{StarGANv2-VC} achieving the highest quality.

Because opinion scores are subjective, we further normalized the MOS to estimate how often volunteers were actually fooled by an RT-DF. In principle, the range of scores given by volunteer $k$ to real audio captures that volunteer’s trust range. Let $\mu_{real}^k$ and $\sigma_{real}^k$ denote the mean and standard deviation of volunteer $k$’s scores on real clips. We estimate that a volunteer would likely be fooled by a clip if they assigned it a score greater than $\mu_{real}^k - \sigma_{real}^k$.

Using this measure, Fig.~\ref{fig:exp1.2} presents the attack success rate for each RT-DF model. We found that \texttt{StarGANv2-VC} had the highest success rate, fooling 46\% of the volunteers. This indicates that although current RT-DF models are not perfect, they can still fool people even under conservative conditions. We note that these results should not be interpreted as the likelihood of a real-world attack succeeding, since our volunteers were actively expecting deepfakes and therefore listened carefully for artifacts. A true victim under pressure would likely overlook at least some of these anomalies.

\textbf{EXP1b.} To analyze the ability of the models to copy identities, we normalized volunteer $k$’s scores on \textit{fake} audio by computing
\[
\frac{score - \mu_{real}^k}{\sigma_{real}^k}.
\]
Figure~\ref{fig:exp2.1} plots the distribution of the normalized scores on fake audio. We observe that the volunteers were mostly indecisive, rating some fake clips as more authentic and some as less. For the majority of cases ($score > -1$), volunteers felt that the identity had been captured well by the top two models.

\paragraph{Video results.}
\textbf{EXP2a.} To assess blind authenticity, we compared the MOS assigned to each identity in both its genuine and deepfake forms. Specifically, for each identity we computed the difference between the mean score given to the fake version and the mean score given to the real version, i.e., $\mu_{fake}^k - \mu_{real}^k$. A positive value indicates that the deepfake identity was perceived as more trustworthy than the genuine one, while a negative value indicates the opposite.

Figure~\ref{fig:exp_1_video} presents the distribution of these values. The distribution is approximately centered around zero and is close to normal, with nearly 50\% of the sample showing a positive difference. In other words, for about half of the identities, the fake video was perceived as more trustworthy than the real one. This suggests that the artifacts or distortions present in current RT-DF videos do not significantly reduce perceived trust and supports our hypothesis that real-time deepfake videos can effectively mislead victims.

\begin{figure}[t]
    \centering
    \includegraphics[width=\columnwidth]{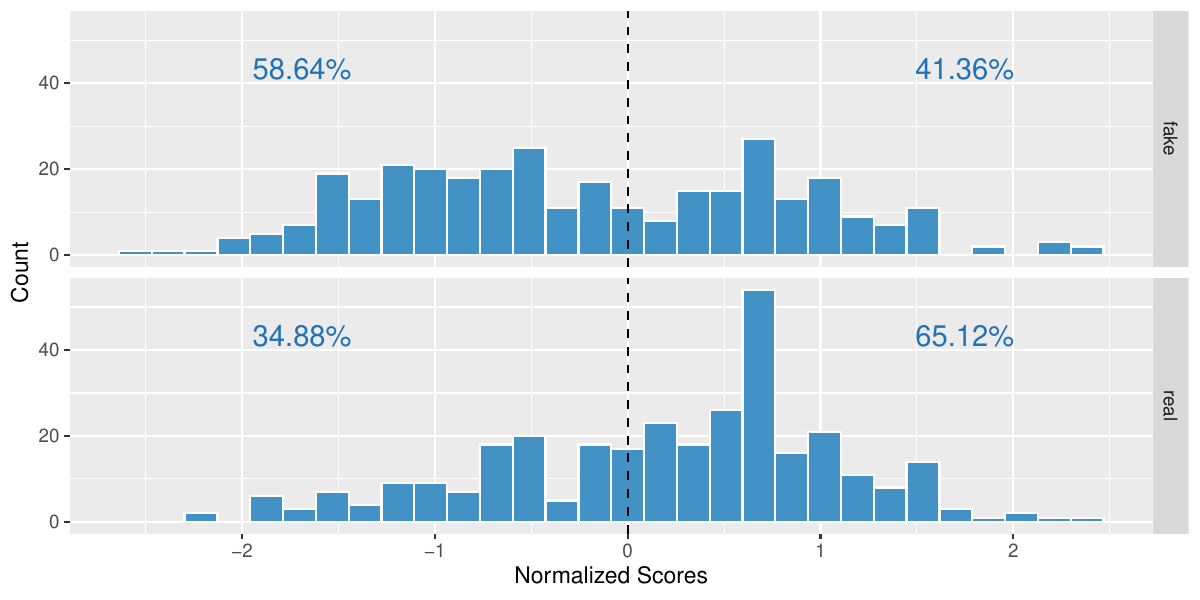}
    \caption{Histogram of real and fake video scores. Positive scores indicate the volunteer believed the video was fake. The graph shows a high false positive rate when participants were asked to detect fakes. There is no significant difference between fake and real scores, suggesting that current deepfake algorithms are challenging for humans to detect accurately.}
    \label{fig:exp_2_video}
\end{figure}

\textbf{EXP2b.} To analyze the detection ability of the volunteers, we calculated the MOS for each user across all videos. Figure~\ref{fig:exp_2_video} compares the resulting distributions for fake and real videos. Two main conclusions emerge. First, participants struggled to accurately identify fake videos, achieving only about 67\% recall. Second, when explicitly asked to look for deepfakes, they became overly suspicious and misidentified genuine videos as fake in approximately 41\% of cases. This high false positive rate indicates a tendency toward over-detection once participants are primed to suspect manipulation.

These findings show that even \textit{after} being informed that some videos are fake, people still cannot reliably distinguish authentic videos from \textit{real-time} deepfake videos. More importantly, the combination of relatively low fake-video detection and high false-positive rates on real videos shows that human judgment alone is not sufficient to defend against this threat.

\textbf{EXP2c.} To analyze the quality (realism) of the video models, we compared the MOS scores of the deepfake models to the MOS of the real videos, again under blind scoring. Figure~\ref{fig:exp3.1} shows the distribution of ratings for each model relative to real video. Based on these results, we selected the best-performing model for the later DF-CAPTCHA evaluation.

\subsection{Takeaways for DF-CAPTCHA}
\begin{figure}[t]
    \centering
    \includegraphics[width=\columnwidth]{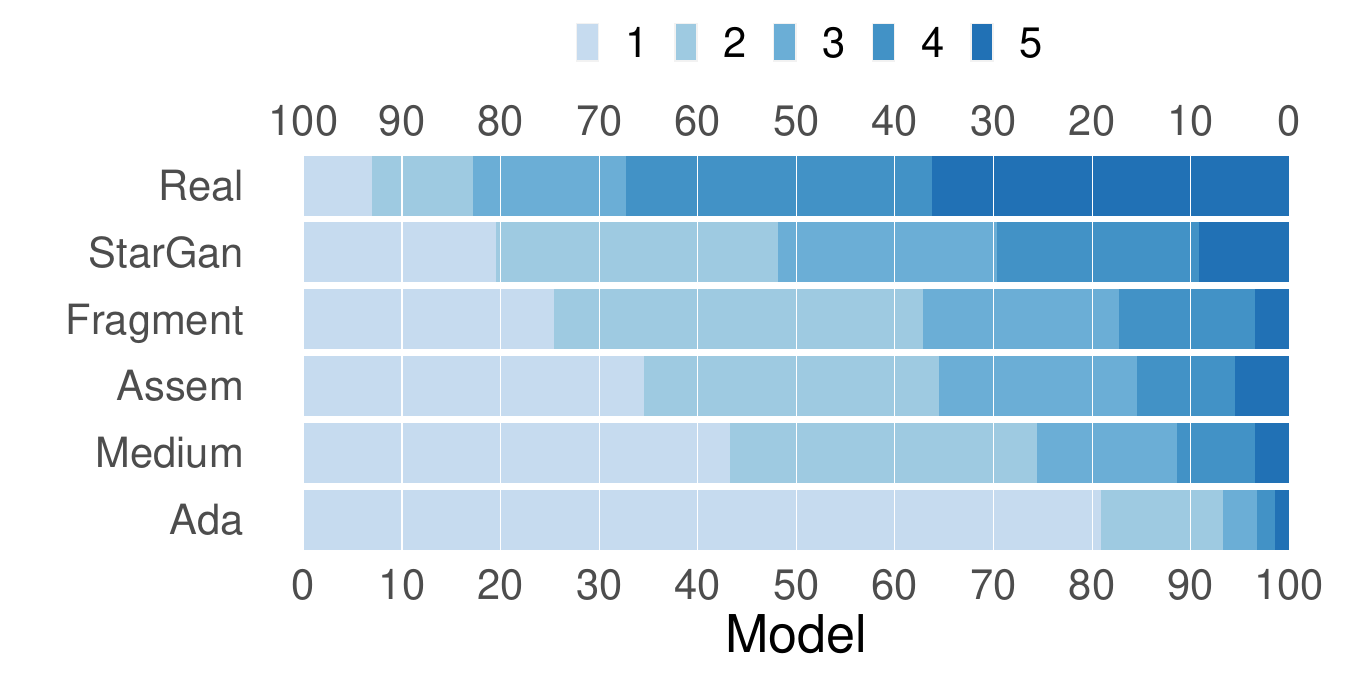}
    \caption{RT-DF Quality - The distribution of ratings which the volunteers gave to each of the RT-DF models and real voice recordings (1: fake, 5: real). }
    \label{fig:exp1.1}
\end{figure}

\begin{figure}[t]
    \centering
    \includegraphics[width=\columnwidth]{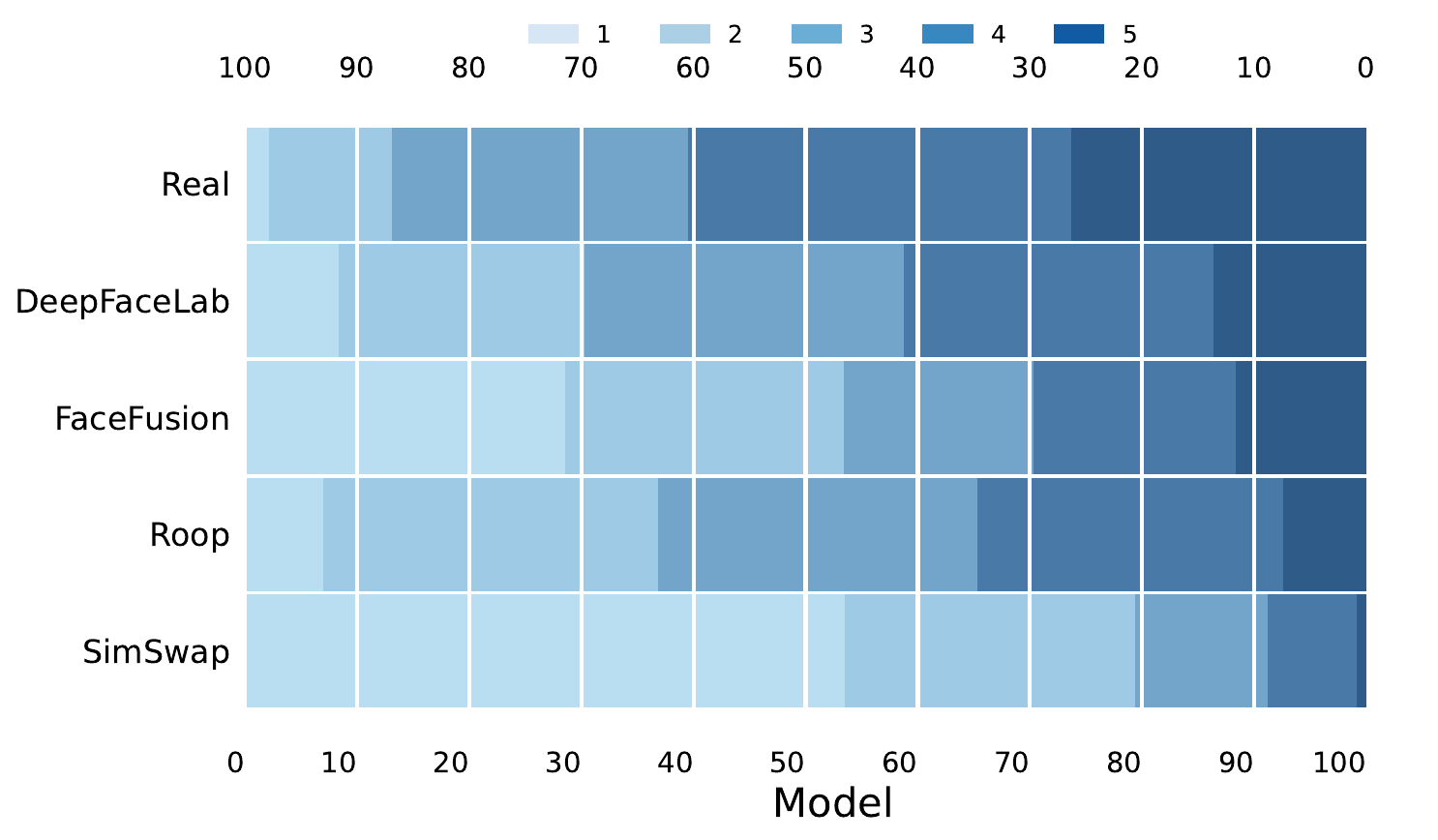}
    \caption{RT-DF Quality - The distribution of ratings which the volunteers gave to each of the DF models and real videos recordings (1: fake, 5: real). }
    \label{fig:exp3.1}
\end{figure}

\begin{figure}[t]
    \centering
    \includegraphics[width=\columnwidth]{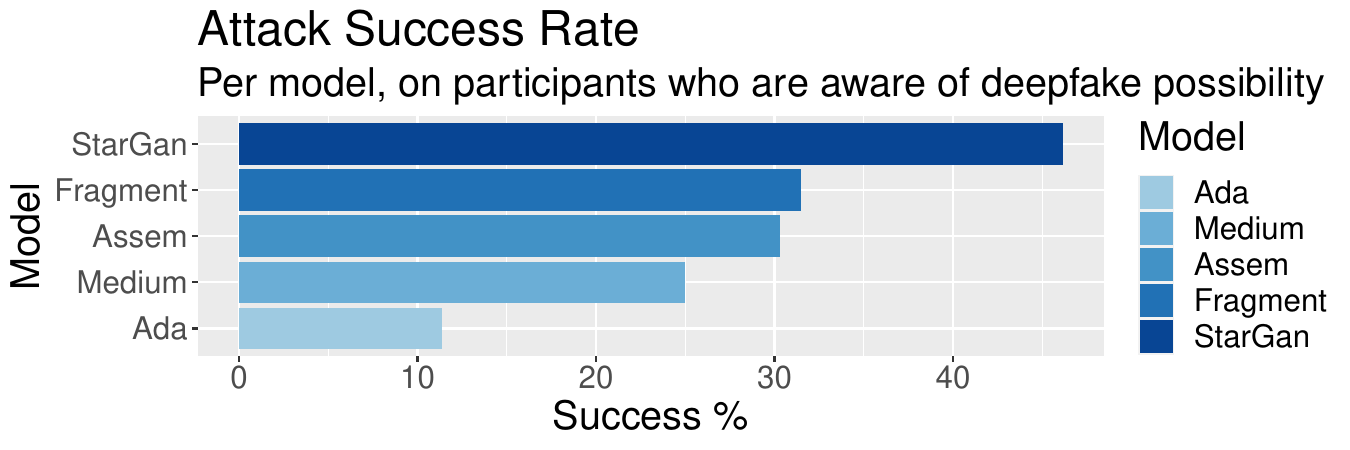}
    \caption{RT-DF Quality - The percent of volunteers fooled by each RT-DF model, even though they were expecting a deepfake.}
    \label{fig:exp1.2}
\end{figure}

\begin{figure}[t]
    \centering
    \includegraphics[width=\columnwidth]{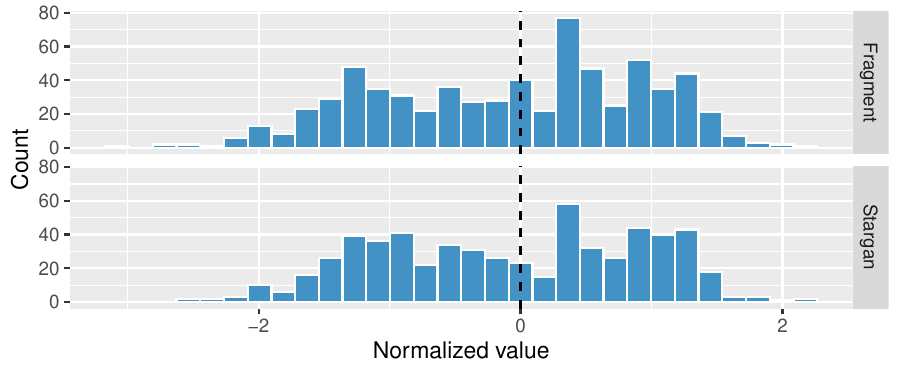}
    \vspace{-1em}
    \caption{RT-DF Identity - A histogram of the normalized MOS scores for how similar RT-DF audio sounds like the target identity $t$. Positive scores are cases where volunteers thought a fake audio sounded more like $t$ than an authentic recording of $t$. }
    \label{fig:exp2.1}
        \vspace{-1em}

\end{figure}

\begin{figure}[t]
    \centering
    \includegraphics[width=\columnwidth]{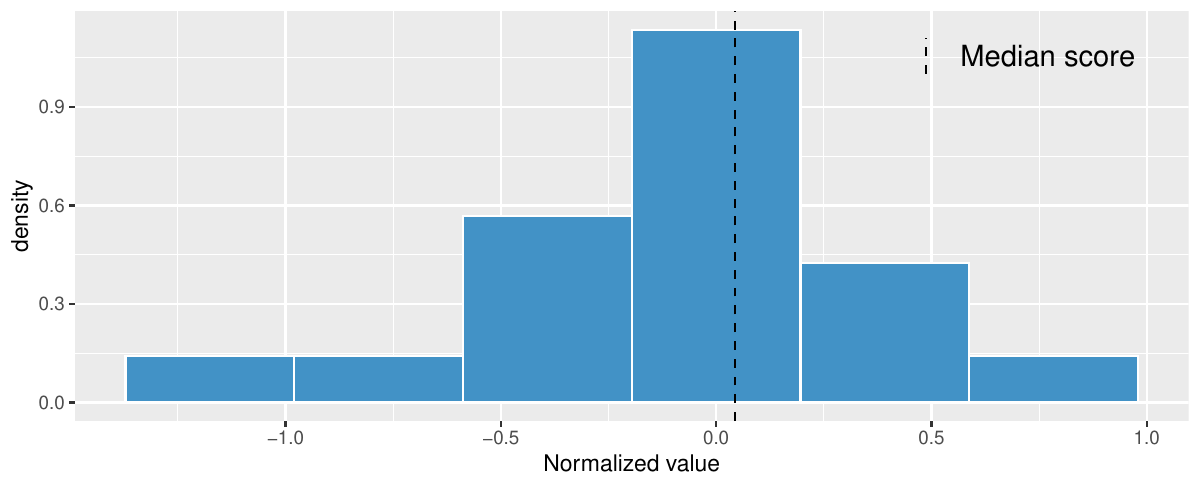}
    \caption{Histogram of the difference between real and fake identity scores in video. A score above 0 indicates the fake identity was perceived as more trustworthy than the real identity. Remarkably, 50\% of fake identities were deemed more trustworthy than their real counterparts. These results demonstrate the significant challenge humans face in detecting deepfakes in videos.}
    \label{fig:exp_1_video}
\end{figure}

Taken together, the audio and video experiments paint a consistent picture. In audio, our original findings showed that even when users are expecting manipulation, modern RT-DF systems can still produce speech that is perceived as realistic and identity-consistent, with \texttt{StarGANv2-VC} fooling 46\% of listeners under these conservative conditions (Figs.~\ref{fig:exp1.1}--\ref{fig:exp2.1}). In video, the new experiments in this extension show that people often trust deepfake videos as much as real ones, and in many cases even more, while also performing poorly when explicitly asked to detect fakes (Figs.~\ref{fig:exp_1_video}--\ref{fig:exp3.1}).

This threat analysis motivates the need for an active defense. Passive perception is unreliable in both modalities, and RT-DF quality is continuing to improve. Therefore, rather than waiting for artifacts to appear naturally, DF-CAPTCHA aims to induce them deliberately by forcing the model outside its stable operating regime.


\label{sec:threat_analysis}
\section{DF-CAPTCHA Evaluation}

We evaluate DF-CAPTCHA across both audio and video modalities, presenting them in parallel to highlight the generality of the active defense principle and to enable direct comparison. The audio pipeline extends our prior conference work~\cite{yasur2023deepfake}, where the core challenge-response framework was first introduced and validated for voice-based RT-DFs. The video pipeline is the primary new contribution of this paper, extending the framework to face-based RT-DFs in video calls. By evaluating both modalities together, we aim to demonstrate that DF-CAPTCHA's advantages are not modality-specific, while also surfacing meaningful differences in how the system behaves across the two domains.

Across both modalities, we report the area under the ROC curve (AUC) and equal error rate (EER). A higher AUC and lower EER both indicate better performance. An AUC of 1.0 indicates a perfect classifier; an AUC of 0.5 indicates random guessing.

\subsection{Experiment Setup}

\subsubsection{Datasets}

\noindent\textbf{Audio.} To evaluate the audio pipeline, we recorded 20 English-speaking volunteers to create both speech and challenge-response datasets. It took each volunteer over an hour to record their data, and all volunteers were compensated for their time. This produced four datasets:

\begin{description}[leftmargin=.3cm]
    \item[$(\mathcal{D}_{real})$] 2,498 samples of real speech (100--250 random sentences per volunteer).
    \item[$(\mathcal{D}_{fake})$] 1,821 samples of RT-DF voice conversion using \texttt{StarGANv2-VC}, the top-performing model from EXP1a. The model was trained to impersonate 6 of the 20 volunteers, augmented with 14 additional voice actors from the VCTK dataset to improve generalization.
    \item[$(\mathcal{D}_{real,r})$] 3,317 samples of real challenge responses across nine tasks. Tasks performed approximately 30 times per volunteer: sing (S), hum tune (HT), coughing (Co), vary volume (V), and talk \& playback (P). Tasks performed approximately 5 times per volunteer: repeat accent (R), clap (Cl), speak with emotion (SE), and vary speed (VS).
    \item[$(\mathcal{D}_{fake,r})$] 16,123 deepfake samples produced by applying \texttt{StarGANv2-VC} to $\mathcal{D}_{real,r}$, excluding same-identity conversions (i.e., where $s = t$).
\end{description}

For all train-test splits, identities were kept disjoint between train and test sets. In addition, the public ASVspoof-DF dataset~\cite{yamagishi2021asvspoof} (22,617 real and 15,000 fake samples) and the RITW dataset~\cite{muller2022does} (19,963 real and 11,816 fake samples) were used to train the realism models. 80\% of the union of these datasets was used for training and 10\% for validation (early stopping).

\vspace{0.5em}
\noindent\textbf{Video.} To evaluate the video pipeline, we recorded 20 volunteers to create both regular video samples and challenge-response videos. Volunteers were compensated for their time. The total corpus spans 310 minutes of footage, averaging 15.5 minutes per volunteer, totaling 20.92 GB of data. This produced four datasets:

\begin{description}[leftmargin=.3cm]
    \item[$(\mathcal{D}_{real})$] 20 samples of regular video per volunteer (105 minutes total; 5.25 minutes per sample).
    \item[$(\mathcal{D}_{fake})$] 20 samples of RT-DF face replacement using \texttt{DeepFaceLab}, the top-performing model from EXP2c, fine-tuned on pairs from the collected dataset.
    \item[$(\mathcal{D}_{real,r})$] 1,000 samples of real challenge responses across ten tasks, each performed approximately 6 times per volunteer: far/close (FC), eyes move (EM), hand occlusion (HO), kiss shape (KS), open mouth (OM), press cheek (PC), puff cheeks (PCs), smile (SM), sunglasses (SG), and turn head (TH).
    \item[$(\mathcal{D}_{fake,r})$] 1,000 deepfake samples produced by applying \texttt{DeepFaceLab} to $\mathcal{D}_{real,r}$.
\end{description}

\subsubsection{Models}

\noindent\textbf{Realism ($\mathcal{R}$).} For the audio pipeline, we evaluated five deepfake detection models. \textbf{SpecRNet}~\cite{kawa2022specrnet} is a lightweight neural architecture inspired by RawNet2~\cite{rawnet2} that achieves competitive detection performance with significantly reduced computational cost. \textbf{One-Class}~\cite{zhang2021one} is based on a ResNet-18 backbone adapted with One-Class Softmax activations for improved generalization. \textbf{GMM-ASVspoof}~\cite{yamagishi2021asvspoof} is a Gaussian mixture model operating on LFCC features, used as the ASVspoof 2021 competition baseline. \textbf{PC-DARTS}~\cite{ge2021raw} is a CNN that automatically learns its own architecture and has shown strong generalization to unseen attacks. \textbf{Local Outlier Factor} (LOF) is a density-based anomaly detector.

For the video pipeline, we evaluated nine deepfake detectors drawn from DeepfakeBench~\cite{DeepfakeBench_YAN_NEURIPS2023}, using their best-performing pretrained weights. \textbf{RECCE}~\cite{cao2022end} uses reconstruction-classification learning to build compact representations of genuine faces. \textbf{FFD}~\cite{dang2020detection} applies attention mechanisms to enhance and localize manipulated facial regions. \textbf{SRM}~\cite{luo2021generalizing} exploits high-frequency image noise through multi-scale feature extraction, residual-guided spatial attention, and cross-modality attention. \textbf{SPSL}~\cite{liu2021spatial} combines spatial image information with phase spectrum features to capture artifacts from cumulative up-sampling. \textbf{Xception}~\cite{rossler2019faceforensics++} is a CNN based on depthwise separable convolutions with residual connections. \textbf{Capsule-Forensics}~\cite{nguyen2019capsule} applies capsule networks to detect a wide range of media forgeries. \textbf{UCF}~\cite{yan2023ucf} disentangles forgery-related features to address overfitting. \textbf{F3Net}~\cite{qian2020thinking} leverages frequency-aware clues via DCT decomposition to handle highly compressed manipulated video. \textbf{CORE}~\cite{ni2022core} regularizes the consistency of representations across augmentations to address overfitting in face forgery detection.

\vspace{0.5em}
\noindent\textbf{Task Verification ($\mathcal{C}$).} For audio, a separate GMM classifier was trained per task on MFCC features, using the ASVspoof 2021 baseline model~\cite{yamagishi2021asvspoof}. Each model classifies between genuine responses to that specific task and all other audio (other tasks or regular speech). A 70-30 train-test split was used. For video, a pretrained ResNet-3D model~\cite{tran2018closer} was fine-tuned to predict which of the ten evaluated tasks is present in a given video clip. A 75-25 train-test split was used.

\vspace{0.5em}
\noindent\textbf{Identity Verification ($\mathcal{I}$).} Both pipelines convert a pretrained recognition model into a zero-shot anomaly detector, using a reference sample $a_t$ captured from the caller before the challenge is issued. This reference is used to verify that the RT-DF is not simply disabled during the challenge.

For audio, we used the ECAPA-TDNN speaker recognition model from the SpeechBrain toolkit~\cite{speechbrain}. Identity consistency is measured as the squared L2 distance between speaker encodings:
\begin{equation}
    \mathcal{I}(a_t, r_c) = \|f^*(a_t) - f^*(r_s)\|^2
\end{equation}
where $f^*$ is the speaker encoding from an inner layer of the model. Smaller scores indicate greater similarity between the pre-challenge voice and the challenge response.

For video, we used a pretrained ResNet-18 model~\cite{he2016deep}. Identity consistency is measured as the cosine similarity between frame encodings:
\begin{equation}
    \mathcal{S}(a_t, r_s) = \frac{f^*(a_t) \cdot f^*(r_s)}{\|f^*(a_t)\| \|f^*(r_s)\|}
\end{equation}
where $f^*$ is the frame encoding from an inner layer of ResNet-18. Higher scores indicate greater similarity. The use of cosine similarity for video (rather than L2 distance for audio) reflects the different embedding geometries of the two pretrained models.

In both cases, we evaluate $\mathcal{I}$ using negative pairings from the same identity $(a_i, r_{c,i})$ and positive pairings from different identities $(a_i, r_{c,j})$, where $a_i, a_j \in \mathcal{D}_{real}$, $r_{c,i}, r_{c,j} \in \mathcal{D}_{real,r}$, and $i \neq j$.

\subsubsection{Experimental Protocol}

For each modality, we conduct four parallel experiments:

\begin{description}[leftmargin=.3cm]
    \item[\textbf{EXP-R} ($\mathcal{R}$)] Comparison of passive deepfake detection (baseline) against active challenge-boosted detection across all realism models.
    \item[\textbf{EXP-C} ($\mathcal{C}$)] Evaluation of the task verification model's ability to confirm that the requested challenge was actually performed.
    \item[\textbf{EXP-I} ($\mathcal{I}$)] Evaluation of the identity verification model's ability to detect when a caller disables the RT-DF during the challenge.
    \item[\textbf{EXP-E} ($\mathcal{R, C, I}$)] End-to-end evaluation of the full DF-CAPTCHA pipeline with all three components operating together.
\end{description}

We do not separately evaluate the timing constraint $\mathcal{T}$ in either modality. It is implemented as a simple binary check (whether the first frame of the response $r_c$ is received within approximately one second of the challenge start) and does not require a learned model.

\subsection{EXP-R: Realism Detection. Active vs. Passive}

The first experiment asks whether issuing a challenge improves the detectability of deepfake artifacts compared to passive detection of ordinary deepfake speech or video.

\noindent\textbf{Audio results.} Table~\ref{tab:eval} reports the AUC and EER for each of the five audio detectors across the baseline scenario and each challenge type. With the exception of SpecRNet, all detectors benefit substantially from the challenge conditions. The best-performing combination overall is GMM-ASVspoof with challenges, reaching an AUC of 0.978 on the talk \& clap (T\&C) task compared to its baseline AUC of 0.949. The LOF detector shows the most dramatic single-task improvement, reaching an AUC of 0.982 on the cough (Co) task from a baseline of 0.678. SpecRNet is the notable exception: its AUC degrades under several challenge conditions (e.g., from 0.952 at baseline to 0.538 on T\&C), suggesting it is not well-suited to the challenge-response setting. In terms of EER, GMM-ASVspoof again performs best with challenges, achieving an EER as low as 0.071 on T\&C compared to its baseline EER of 0.122.

\input{eval_table.tex}

\begin{figure}[t]
    \centering
    \includegraphics[width=0.6\columnwidth]{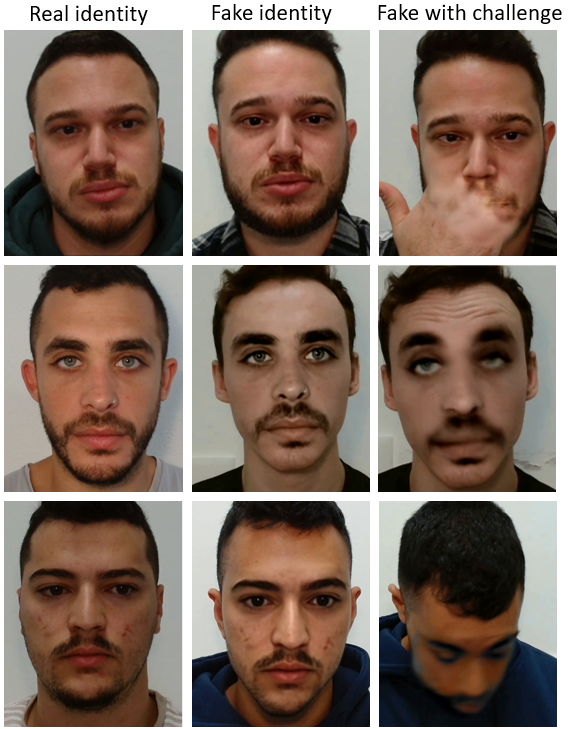}
    \caption{A sample of video challenges showing their effect on stressing the limitations of modern deepfake pipelines.}
    \label{fig:df_exp_challanges}
\end{figure}

\noindent\textbf{Video results.} Table \ref{tab:auc_eer_video} reports the AUC and EER for each of the nine video detectors. Every model shows improved performance when challenges are introduced without exception because the challenges push the pipelines our of their capabilities causing large artifacts, as shown in Fig. \ref{fig:df_exp_challanges}. CORE demonstrates the most dramatic enhancement, improving from a baseline AUC of 0.777 to a peak of 0.987 on the eye movement (EM) task, and achieving an EER as low as 0.04 on the same task from a baseline EER of 0.40. SRM is the strongest model on the hand occlusion (HO) task (AUC 0.974, EER 0.09) and performs consistently well across tasks. UCF is the weakest overall, with several challenge conditions failing to improve substantially over its baseline (e.g., AUC of 0.610 on HO vs.\ a baseline of 0.694), though it does benefit on some tasks such as EM (AUC 0.880). Xception achieves a near-perfect AUC of 0.993 on the EM task.

\input{eval_table_mean_eer.tex}

\noindent\textbf{Shared insight.} The pattern is consistent across both modalities: challenges force RT-DF models to operate outside their training distribution, amplifying artifacts that are otherwise subtle or absent in ordinary deepfake content. The improvement is broad, affecting most models in both domains, and the best passive baseline in each modality is substantially weaker than the best challenge-boosted result. This validates the core premise of DF-CAPTCHA: active elicitation is more reliable than passive observation.

\subsection{EXP-C: Task Verification}

The second experiment evaluates whether the system can confirm that the caller actually performed the requested challenge, as opposed to staying silent, performing a different action, or producing a deepfake response that fails to replicate the requested task.

\begin{figure}[t]
    \centering
    \includegraphics[width=0.9\columnwidth]{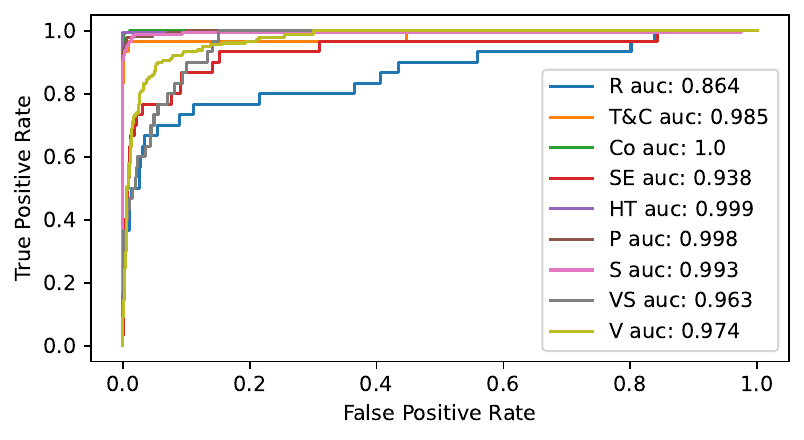}
    \caption{The performance of the audio task detection model $\mathcal{C}$.}
    \label{fig:exp2b}
\end{figure}

\noindent\textbf{Audio results.} Figure~\ref{fig:exp2b} shows the AUC of the audio task detection model $\mathcal{C}$ across all nine tasks. The model performs well across most tasks, with several reaching near-perfect AUC: cough (Co) achieves an AUC of 1.0, hum tune (HT) achieves 0.999, playback (P) achieves 0.998, and sing (S) achieves 0.993. The weakest performer is the repeat accent (R) task, which achieves an AUC of only 0.864. This can be attributed to two compounding factors: the subtle acoustic differences between genuine and imitated accent modifications are inherently difficult to distinguish, and many volunteers struggled to produce convincing accent variation, resulting in responses that closely resembled normal speech and were therefore hard to classify.

\begin{figure}[t]
    \centering
    \includegraphics[width=0.9\columnwidth]{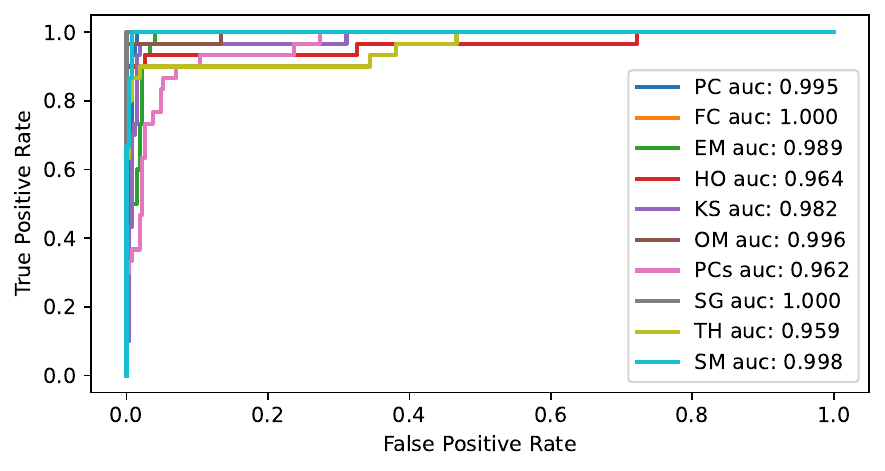}
    \caption{The performance of the video task detection model $\mathcal{C}$.}
    \label{fig:exp2b_vid}
\end{figure}

\noindent\textbf{Video results.} Figure~\ref{fig:exp2b_vid} shows the AUC of the video task detection model $\mathcal{C}$ across all ten tasks. Performance is near-perfect across the board. Far/close (FC) and sunglasses (SG) both achieve an AUC of 1.000. Kiss shape (KS) achieves 0.982, open mouth (OM) achieves 0.996, press cheek (PC) achieves 0.995, eyes move (EM) achieves 0.989, puff cheeks (PCs) achieves 0.962, smile (SM) achieves 0.998, hand occlusion (HO) achieves 0.964, and turn head (TH) achieves 0.959. Even when adversaries attempt to evade by remaining motionless, or when the deepfake model fails to synthesize the required movement, $\mathcal{C}$ maintains strong detection efficacy. Tasks requiring coordinated and complex facial deformation, particularly puff cheeks and open mouth, are especially effective, likely because these movements impose particularly demanding constraints on real-time face manipulation models.

\noindent\textbf{Shared insight.} Task verification is highly reliable in both modalities, and in both cases the weaker tasks are those that are either acoustically or visually ambiguous, or that humans themselves find difficult to perform consistently. This points to a key design principle: when selecting challenges for DF-CAPTCHA, it is important to favor tasks that are both easy for genuine callers to perform and sufficiently distinct from baseline behavior to be robustly classifiable. In the video domain, complex facial deformations appear to satisfy both criteria particularly well.

\subsection{EXP-I: Identity Verification}

The third experiment evaluates whether the system can detect a simple but important evasion strategy: the attacker disables the RT-DF during the challenge and responds as their real self, then re-enables the deepfake afterward.

\begin{figure}[t]
    \centering
    \includegraphics[width=0.9\columnwidth]{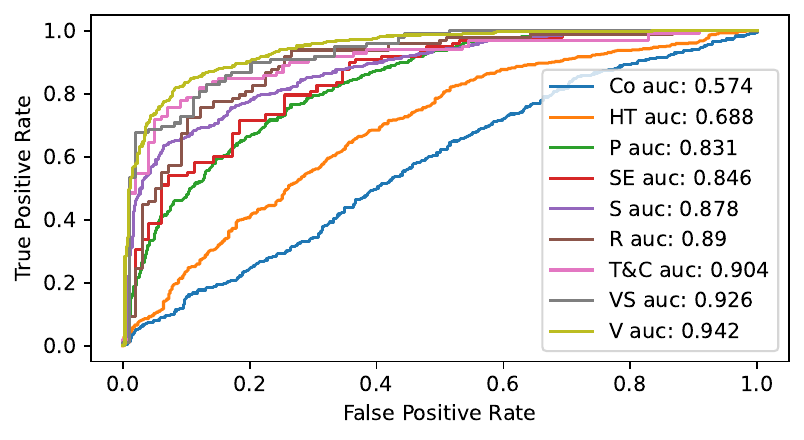}
    \caption{The performance of the unsupervised audio identity detection model $\mathcal{I}$ for different tasks.}
    \label{fig:exp2c}
\end{figure}

\noindent\textbf{Audio results.} Figure~\ref{fig:exp2c} shows the AUC of the audio identity model $\mathcal{I}$ across all nine tasks. Performance varies considerably by task. The strongest results are achieved on tasks that preserve a clear speaker signature: vary volume (V) achieves an AUC of 0.942, vary speed (VS) achieves 0.926, talk \& clap (T\&C) achieves 0.904, and repeat accent (R) achieves 0.890. The weakest tasks are cough (Co) with an AUC of 0.574 and hum tune (HT) with an AUC of 0.688. This is expected: humming and coughing suppress or distort the spectral characteristics that distinguish speakers, providing insufficient identity signal for the model to work with.

\begin{figure}[t]
    \centering
    \includegraphics[width=0.9\columnwidth]{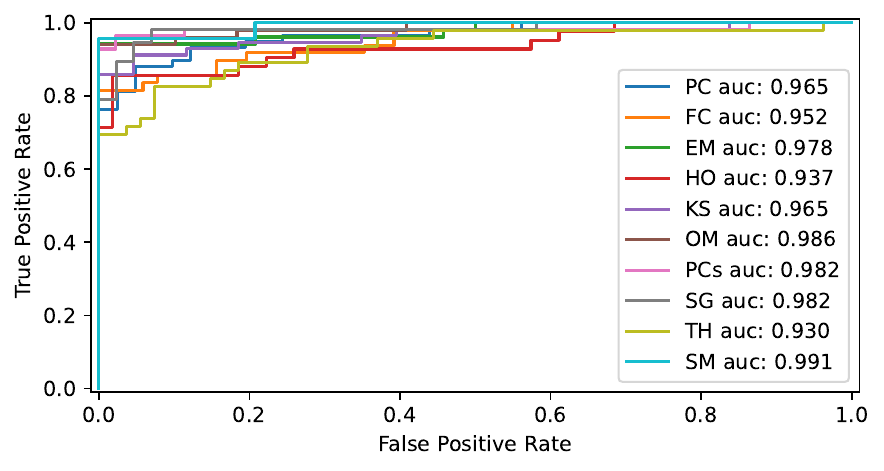}
    \caption{The performance of the unsupervised video identity detection model $\mathcal{I}$ for different tasks.}
    \label{fig:exp2c_vid}
\end{figure}

\noindent\textbf{Video results.} Figure~\ref{fig:exp2c_vid} shows the AUC of the video identity model $\mathcal{I}$ across all ten tasks. Overall performance is strong and substantially more uniform than in the audio domain. The best results are achieved on smile (SM) at AUC 0.991, open mouth (OM) at 0.986, puff cheeks (PCs) at 0.982, sunglasses (SG) at 0.982, and eyes move (EM) at 0.978. The most challenging task is turn head (TH), which achieves an AUC of 0.930, which is still strong but the lowest across the ten tasks. This is because turning the head temporarily occludes discriminative facial features for several frames, reducing the reliability of frame-level identity encoding. Hand occlusion (HO) achieves 0.937 for similar reasons, with the hand partially covering the face during execution.

\noindent\textbf{Shared insight.} The identity verifier performs well in both modalities when the challenge preserves sufficient identity signal, but reveals modality-specific weaknesses. In audio, tasks that suppress speaker-specific spectral content (humming, coughing) degrade performance significantly. In video, tasks that occlude the face (turning the head, hand occlusion) pose the most difficulty, though the degradation is milder than in the audio case. In both domains, the practical implication is the same: DF-CAPTCHA should either avoid tasks that are known to obscure identity, or compensate by requiring the caller to complete additional tasks in sequence that collectively preserve enough identity signal for reliable verification.

\subsection{EXP-E: End-to-End System Performance}

The final experiment evaluates the complete DF-CAPTCHA pipeline, with all three components ($\mathcal{R}$, $\mathcal{C}$, and $\mathcal{I}$) operating in tandem. For each model, thresholds are set such that the false positive rate is FPR~$= 0.01$. This conservative setting prioritizes minimizing disruption to legitimate callers. The compounding effects of each model's individual successes and failures are reflected in the end-to-end metrics.

\begin{figure}[t]
    \centering
    \includegraphics[width=\columnwidth]{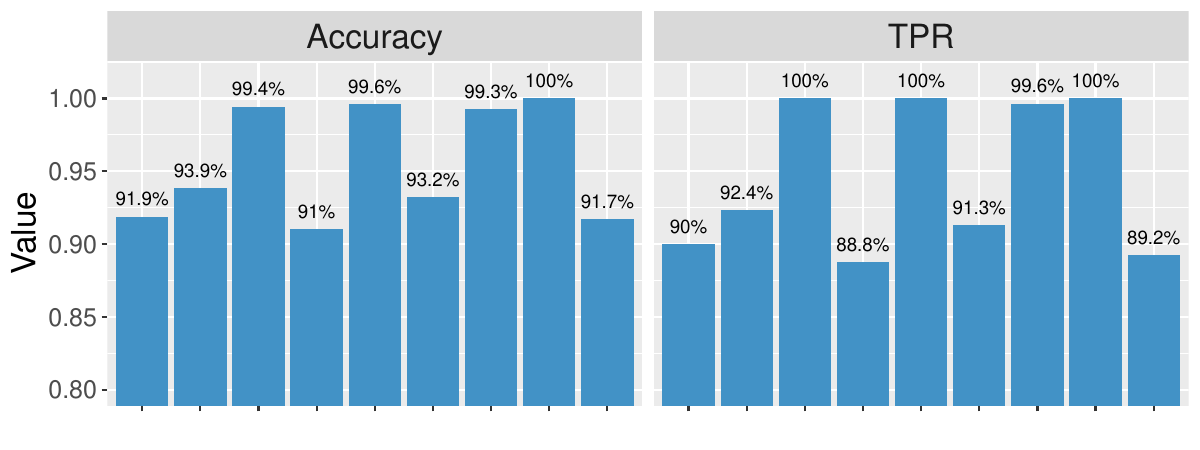}
    \includegraphics[width=\columnwidth]{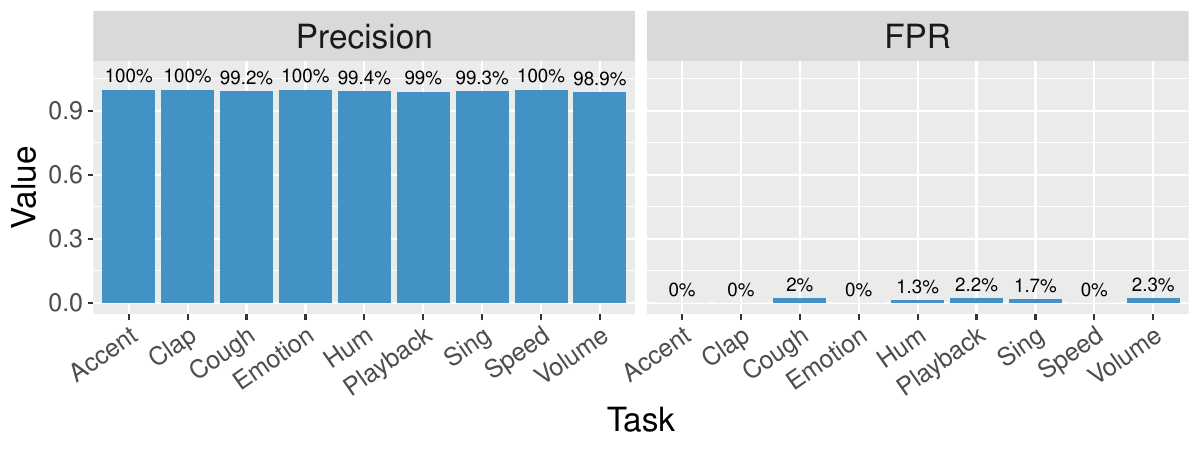}
    \caption{The end-to-end performance of the DF-CAPTCHA system on audio (voice).}
    \label{fig:exp2d}
\end{figure}

\noindent\textbf{Audio results.} We passed 3,317 real responses and 8,758 deepfake responses through the full pipeline. As shown in Figure~\ref{fig:exp2d}, the system achieves a TPR of 0.89--1.00 depending on the selected task, with FPR ranging from 0.0 to 2.3\% and overall accuracy of 91--100\%. The best-performing tasks include cough (Co) and sing (S), both reaching accuracy of 99--100\% with TPR at or near 1.00. The weakest task is repeat accent (R), reflecting the lower performance of the task detector $\mathcal{C}$ on this challenge identified in EXP-C. By comparison, the best passive baseline (SpecRNet) achieves a TPR of only 0.66 and an accuracy of 71\% at the same FPR constraint of 0.01. DF-CAPTCHA therefore represents a substantial improvement of approximately 20--29 percentage points in accuracy over the best passive method.

\begin{figure}[t]
    \centering
    \includegraphics[width=\columnwidth]{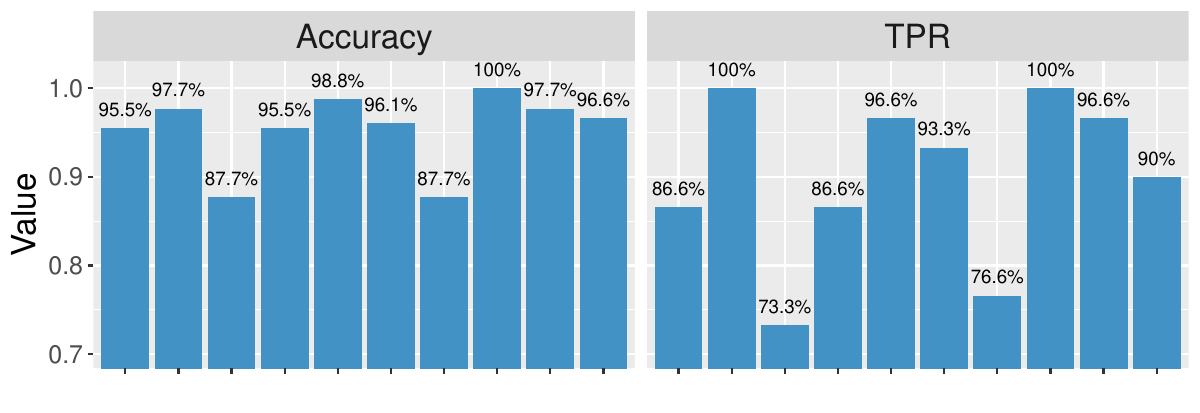}
    \includegraphics[width=\columnwidth]{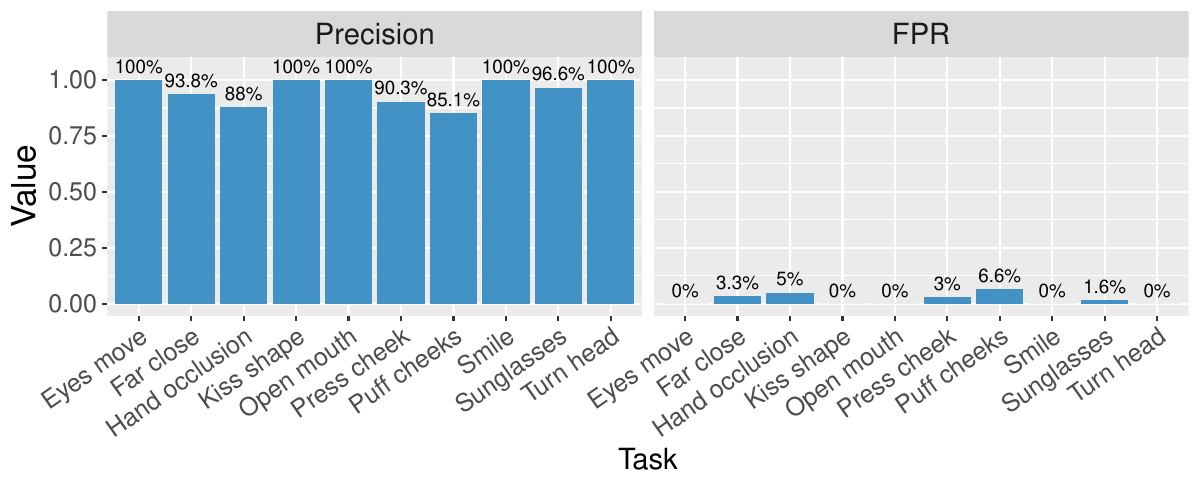}
    \caption{The performance of the ensure DF-CAPTCHA system in video domain (end-to-end).}

    \label{fig:exp2d_vid}
\end{figure}

\noindent\textbf{Video results.} We passed 3,317 genuine responses and 8,758 deepfake responses through the full video pipeline. As shown in Figure~\ref{fig:exp2d_vid}, the system achieves a TPR of 0.73--1.00 depending on the selected challenge, with FPR ranging from 0.0 to 6\% and overall accuracy of 87--100\%. The best-performing tasks are open mouth (OM) and smile (SM), both reaching accuracy of 98--100\% with TPR at or near 1.00. The weakest task is puff cheeks (PCs), which achieves a TPR of 0.733 and accuracy of 87.7\% which is still well above the passive baseline, but reflects that some challenge types place more simultaneous pressure on $\mathcal{I}$ and $\mathcal{R}$ together. The best passive baseline (CORE) achieves a TPR of only 0.55 and an accuracy of 75\% under the same FPR constraint. DF-CAPTCHA therefore achieves an improvement of 12--25 percentage points in accuracy over the best passive method, depending on the task.

\noindent\textbf{Cross-modal comparison.} The audio pipeline achieves a higher floor on TPR (0.89 vs.\ 0.73), while the video pipeline exhibits a slightly wider FPR range under the same threshold setting, reflecting greater variance across video challenge types in the identity and realism components. In both cases, however, the improvement over passive baselines is large and consistent: DF-CAPTCHA outperforms the best passive detector by 20--29 percentage points in audio and 12--25 percentage points in video. The best-performing tasks in each modality reach near-perfect accuracy, and the combination of $\mathcal{R}$, $\mathcal{C}$, and $\mathcal{I}$ is more powerful than any single component or any passive detector operating alone. These results confirm that the active challenge-response principle generalizes robustly across both modalities.

\label{sec:evaluation}

\section{Discussion}
\subsection{Discussion of the Results}
The results confirm that active deepfake detection substantially outperforms passive
detection alone. By requiring callers to perform unpredictable challenges, DF-CAPTCHA
forces real-time deepfake pipelines outside their stable operating regime, amplifying
artifacts that realism, task, and identity verifiers can exploit. Crucially, this
benefit is not architecture-specific: multiple realism detectors improved under
challenge conditions, indicating that the framework exposes general failure modes
rather than detector-specific weaknesses. This robustness makes DF-CAPTCHA well-suited
to practical deployments such as remote interviews, call centers, and online meetings,
where conditions and attack pipelines vary.

Several limitations remain. The framework targets real-time attacks and does not address
pre-recorded deepfake media. As generative models improve, some challenges may lose
effectiveness and require replacement. The protocol can introduce user friction if
deployed too aggressively in low-risk settings, and false positives remain a concern
when legitimate users perform challenges under poor recording conditions.

Future work should expand the challenge library to stay ahead of generative advances,
evaluate real-world usability (latency, accessibility, challenge selection policy), and
improve robustness against adaptive adversaries by developing more challenges according to the present day's deepfake pipeline limitations.

\subsection{XAI in Video CAPTCHAs}
\begin{figure}[h]
    \centering
    \includegraphics[width=0.6\columnwidth]{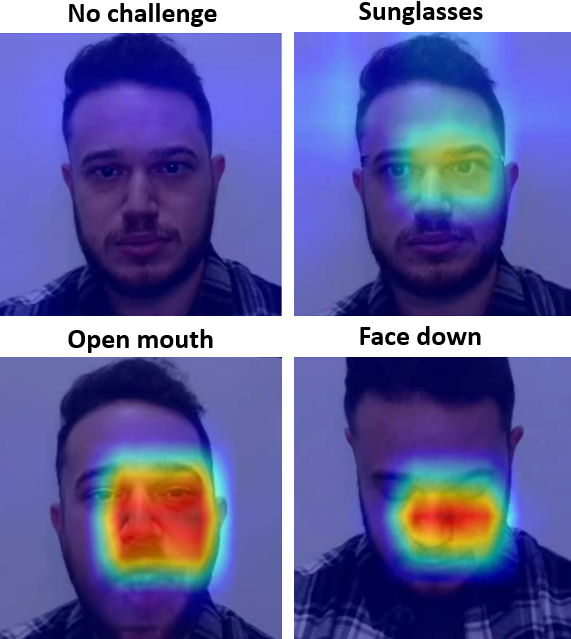}
    \caption{XAI in Video CAPTCHA Analysis. Attention of a face neural network for three CAPTCHA types vs. no CAPTCHA. Higher activation observed in frames with active challenges.}
    \label{fig:XAI}
\end{figure}

To better understand why challenge-based detection improves performance, we used explainable AI (XAI) techniques to analyze the behavior of a deepfake detection network when processing fake videos with and without the proposed CAPTCHA challenges (illustrated in Fig. \ref{fig:XAI}). Using the FaceForensics++-based model in \cite{rossler2019faceforensics++}, we observed substantially stronger activation patterns in frames that contained an active challenge than in comparable frames without a challenge. This confirms that the challenges expose visual irregularities that the detector identifies as salient evidence of manipulation.

More specifically, frames containing an executed challenge were assigned a substantially higher probability of being fake. This trend was consistent across multiple challenge types, indicating that the effect is not limited to a single gesture or motion pattern. The results therefore provide interpretability-based support for our main hypothesis: dynamic, challenge-induced behaviors make real-time deepfakes more detectable by revealing anomalies that may remain weak or ambiguous during ordinary conversation.

The XAI analysis also helps clarify what the detector is attending to. The most salient regions were typically located around the challenged facial or motion-related areas, suggesting that these induced behaviors create localized inconsistencies in geometry, appearance, or temporal coherence. This finding is important for two reasons. First, it provides a qualitative explanation for the quantitative gains observed in the evaluation. Second, it offers a practical guide for refining future challenges, since highly salient behaviors are likely to be particularly useful for active detection. Overall, the XAI results strengthen the argument that DF-CAPTCHA improves detection not merely by adding interaction, but by deliberately eliciting behaviors that expose structural weaknesses in current real-time deepfake pipelines.

\section{Conclusion}
In this paper, we presented DF-CAPTCHA, an active defense against real-time deepfake impersonation attacks in live communications. Rather than relying on passive artifact detection alone, our method challenges callers with simple tasks that are easy for humans but difficult for current real-time deepfake systems to perform convincingly. Our results show that this challenge-response approach substantially improves detection in both audio and video settings, while also addressing the growing limitations of human judgment in identifying deepfakes.

A central contribution is our expansion of DF-CAPTCHA from audio to video, enabling a broader defense against multimodal real-time impersonation. Through new threat analysis, user studies, and experiments on face-based real-time deepfakes, we show that the same active-defense principle generalizes effectively beyond voice cloning to video manipulation. Overall, DF-CAPTCHA offers a practical, extensible, and forward-looking framework for securing digital interactions against next-generation social engineering attacks.

% \appendix

% \section{Sample Appendix Section}
% \label{sec:sample:appendix}

\bibliographystyle{IEEEtran}
\bibliography{cas-refs}

\section*{Acknowledgments}
This work was supported by the Zuckerman STEM Leadership Program. Some figures were made with icograms.com. ChatGPT 5.2 has been used to improve the English quality of this paper and all edits have been reviewed by the authors.

\section{Biography Section}
\begin{footnotesize}

\noindent\textbf{Guy Frankovits} is a researcher and master's degree student in the
Department of Software and Information Systems Engineering at Ben-Gurion University
of the Negev. His research interests include machine learning, deep learning, and
deepfake detection in both audio and video.

\smallskip

\noindent\textbf{Lior Yasur} is a researcher and master's degree student in the
Department of Software and Information Systems Engineering at Ben-Gurion University
of the Negev. His research interests include machine learning and adversarial learning.

\smallskip

\noindent\textbf{Fred Grabovski} is a researcher and master's degree student in the
Department of Software and Information Systems Engineering at Ben-Gurion University
of the Negev. His research interests include machine learning and adversarial learning.

\smallskip

\noindent\textbf{Yisroel Mirsky} is a tenured Assistant Professor and Zuckerman
Faculty Scholar at Ben-Gurion University, where he leads the Offensive AI Research
Lab. He received his Ph.D. from BGU in 2018 and completed postdoctoral training at
the Georgia Institute of Technology under Prof.\ Wenke Lee. His research spans AI
safety, AI agent security, deepfakes, and adversarial machine learning. Dr.\ Mirsky
is an ERC-STG 2025 recipient and serves on the board of COSAI (Coalition for Secure
AI), an industry consortium whose members include Google, Microsoft, Amazon, and
Intel. He has published at leading security venues including USENIX, CCS, NDSS,
Black Hat, and DEF CON. His work has been covered by Wired, The Wall Street Journal,
Forbes, and Scientific American, and has earned over \$10,000 in bug bounties.
Notable projects include exposing vulnerabilities in the US 911 emergency services
and revealing the threat of deepfakes in medical imaging, both covered by
The Washington Post.

\end{footnotesize}

\vfill

\end{document}

%% file: captcha_table_all.tex
% \usepackage{graphicx}
\begin{table*}[]
\caption{Examples of video-based (top) and audio-based (bottom) tasks which can be used as challenges in a D-CAPTCHA. Strong challenges are hard for the adversary on all four constraints: realism, identity, complexity and time.}\label{tab:captchas_combined}

\resizebox{\textwidth}{!}{%
\def\arraystretch{0.85}%

\begin{tabular}{lccccccccc}
\multicolumn{3}{r|}{}                                                                          & \multicolumn{4}{c|}{Hardness}                                                                   & \multicolumn{1}{c|}{Weakness}          & \multicolumn{2}{c|}{Effectiveness}                                            \\
\multicolumn{1}{c}{\textbf{Task ($T$)}} & \textbf{Acronym} & \multicolumn{1}{c|}{\textbf{Usability}} & \textbf{Realism} & \textbf{Identity} & \textbf{Task} & \multicolumn{1}{c|}{\textbf{Time}} & \multicolumn{1}{c|}{\textbf{Evasions}} & \textbf{Naive   Attacker} & \multicolumn{1}{c|}{\textbf{Advanced   Attacker}} \\ \hline\hline
\multicolumn{10}{l}{\textit{Video-based Tasks}} \\ \hline
\textit{Drop   Object}            & DO               & \multicolumn{1}{c|}{$\bullet$}          & $\bullet$        & $\bullet$           & $\circ$           & \multicolumn{1}{c|}{$\bullet$}     & \multicolumn{1}{c|}{}                  & $\bullet$                 & \multicolumn{1}{c|}{$\bullet$}                      \\
\textit{Bounce   Object}          & BO               & \multicolumn{1}{c|}{$\bullet$}          & $\bullet$        & $\bullet$           & $\circ$           & \multicolumn{1}{c|}{$\bullet$}     & \multicolumn{1}{c|}{}                  & $\bullet$                 & \multicolumn{1}{c|}{$\bullet$}                    \\
\textit{Fold Shirt}               & FS               & \multicolumn{1}{c|}{$\bullet$}          & $\bullet$        & $\bullet$           & $\bullet$           & \multicolumn{1}{c|}{$\bullet$}     & \multicolumn{1}{c|}{}                & $\bullet$                 & \multicolumn{1}{c|}{$\circ$}                    \\
\textit{Stroke hair}              & SH               & \multicolumn{1}{c|}{$\bullet$}          & $\bullet$        & $\bullet$           & $\circ$           & \multicolumn{1}{c|}{$\bullet$}     & \multicolumn{1}{c|}{}                  & $\bullet$                 & \multicolumn{1}{c|}{$\bullet$}                    \\
\textit{Interact with Background} & IB               & \multicolumn{1}{c|}{$\circ$}            & $\bullet$        & $\circ$             & $\circ$             & \multicolumn{1}{c|}{$\bullet$}     & \multicolumn{1}{c|}{}                & $\circ$                   & \multicolumn{1}{c|}{$\circ$}                      \\
\textit{Spill Water}              & SW               & \multicolumn{1}{c|}{$\circ$}            & $\bullet$        & $\bullet$           & $-$             & \multicolumn{1}{c|}{$\bullet$}     & \multicolumn{1}{c|}{}                    & $\circ$                   & \multicolumn{1}{c|}{$\bullet$}                      \\
\textit{Pick Up Object}           & PO               & \multicolumn{1}{c|}{$\bullet$}          & $\bullet$        & $\bullet$           & $\circ$           & \multicolumn{1}{c|}{$\bullet$}     & \multicolumn{1}{c|}{}                  & $\bullet$                 & \multicolumn{1}{c|}{$\circ$}                    \\
\textit{Hand Expressions}         & HE               & \multicolumn{1}{c|}{$\bullet$}          & $\bullet$        & $\bullet$           & $\circ$             & \multicolumn{1}{c|}{$\bullet$}     & \multicolumn{1}{c|}{}                & $\bullet$                 & \multicolumn{1}{c|}{$\bullet$}                    \\
\textit{Tongue Motion}            & TM               & \multicolumn{1}{c|}{$\circ$}            & $\bullet$        & $\bullet$           & $-$           & \multicolumn{1}{c|}{$\bullet$}     & \multicolumn{1}{c|}{}                      & $\bullet$                 & \multicolumn{1}{c|}{$\bullet$}                    \\
\textit{Fold Ear}                 & FE               & \multicolumn{1}{c|}{$\bullet$}          & $\bullet$        & $\bullet$           & $\circ$           & \multicolumn{1}{c|}{$\bullet$}     & \multicolumn{1}{c|}{}                  & $\bullet$                 & \multicolumn{1}{c|}{$\bullet$}                          \\
\textit{Face Occlusions}          & FO               & \multicolumn{1}{c|}{$\circ$}            & $\bullet$        & $\circ$             & $\bullet$           & \multicolumn{1}{c|}{$\bullet$}     & \multicolumn{1}{c|}{}                & $\bullet$                 & \multicolumn{1}{c|}{$\bullet$}                          \\
\textit{Remove Glasses}           & RG               & \multicolumn{1}{c|}{$\circ$}            & $\circ$          & $\bullet$           & $\bullet$           & \multicolumn{1}{c|}{$\bullet$}     & \multicolumn{1}{c|}{}                & $\bullet$                 & \multicolumn{1}{c|}{$\circ$}                          \\
\textit{Walking Out of Frame}     & WOF              & \multicolumn{1}{c|}{$\bullet$}          & $\circ$          & $-$                 & $\bullet$           & \multicolumn{1}{c|}{$\bullet$}     & \multicolumn{1}{c|}{bypass}          & $\circ$                   & \multicolumn{1}{c|}{$-$}                          \\
\textit{Moving Close and Far}     & MCF              & \multicolumn{1}{c|}{$\bullet$}          & $\bullet$        & $\bullet$           & $\bullet$           & \multicolumn{1}{c|}{$\bullet$}     & \multicolumn{1}{c|}{}                & $\bullet$                 & \multicolumn{1}{c|}{$\bullet$}                          \\
\textit{Turn Head}                & TH               & \multicolumn{1}{c|}{$\bullet$}          & $\bullet$        & $\circ$             & $\bullet$           & \multicolumn{1}{c|}{$\bullet$}     & \multicolumn{1}{c|}{}                & $\bullet$                 & \multicolumn{1}{c|}{$\bullet$}                          \\
\textit{Press Nose}               & PN               & \multicolumn{1}{c|}{$\bullet$}          & $\bullet$        & $\bullet$           & $\circ$           & \multicolumn{1}{c|}{$\bullet$}     & \multicolumn{1}{c|}{}                  & $\bullet$                 & \multicolumn{1}{c|}{$\bullet$}                          \\
\textit{Pressing Cheek}           & PC               & \multicolumn{1}{c|}{$\bullet$}          & $\bullet$        & $\bullet$           & $\circ$           & \multicolumn{1}{c|}{$\bullet$}     & \multicolumn{1}{c|}{}                  & $\bullet$                 & \multicolumn{1}{c|}{$\bullet$}                          \\
\textit{Show Teeth}               & ST               & \multicolumn{1}{c|}{$\bullet$}          & $\circ$          & $\bullet$           & $\circ$           & \multicolumn{1}{c|}{$\bullet$}     & \multicolumn{1}{c|}{}                  & $\bullet$                 & \multicolumn{1}{c|}{$\circ$}                          \\
\textit{Toss Coin}                & TC               & \multicolumn{1}{c|}{$\circ$}            & $\circ$          & $\circ$             & $-$           & \multicolumn{1}{c|}{$\bullet$}     & \multicolumn{1}{c|}{}                      & $\bullet$                 & \multicolumn{1}{c|}{$\bullet$}                          \\
\textit{Shine Light on Face}      & SF               & \multicolumn{1}{c|}{$\circ$}            & $\circ$          & $\bullet$           & $\bullet$           & \multicolumn{1}{c|}{$\bullet$}     & \multicolumn{1}{c|}{bypass}          & $\bullet$                 & \multicolumn{1}{c|}{$\circ$}                          \\
\textit{Drum on Table}            & DT               & \multicolumn{1}{c|}{$\bullet$}          & $\circ$          & $\bullet$           & $\bullet$           & \multicolumn{1}{c|}{$\bullet$}     & \multicolumn{1}{c|}{}                & $\bullet$                 & \multicolumn{1}{c|}{$\circ$}                          \\
\textit{Display Items in Hand}    & DI               & \multicolumn{1}{c|}{$\bullet$}          & $\circ$          & $\bullet$           & $\circ$           & \multicolumn{1}{c|}{$\bullet$}     & \multicolumn{1}{c|}{}                  & $\bullet$                 & \multicolumn{1}{c|}{$\circ$}                          \\
\textit{Turn Around}              & TA               & \multicolumn{1}{c|}{$\bullet$}          & $\bullet$        & $-$                 & $\bullet$           & \multicolumn{1}{c|}{$\bullet$}     & \multicolumn{1}{c|}{bypass}          & $\bullet$                 & \multicolumn{1}{c|}{$\bullet$}                          \\
\textit{Roll Head}                & RH               & \multicolumn{1}{c|}{$\bullet$}          & $\bullet$        & $\bullet$           & $\circ$           & \multicolumn{1}{c|}{$\bullet$}     & \multicolumn{1}{c|}{}                  & $\bullet$                 & \multicolumn{1}{c|}{$\bullet$}                          \\
\textit{Puff Cheeks}              & PC               & \multicolumn{1}{c|}{$\bullet$}          & $\bullet$        & $\bullet$           & $\circ$           & \multicolumn{1}{c|}{$\bullet$}     & \multicolumn{1}{c|}{}                  & $\bullet$                 & \multicolumn{1}{c|}{$\bullet$}                          \\
\textit{Mouth Shapes}             & MS               & \multicolumn{1}{c|}{$\circ$}            & $\circ$          & $\bullet$           & $\circ$           & \multicolumn{1}{c|}{$\bullet$}     & \multicolumn{1}{c|}{}                  & $\bullet$                 & \multicolumn{1}{c|}{$\bullet$}                    \\
\textit{Eye Gestures}             & VS               & \multicolumn{1}{c|}{$\circ$}            & $\circ$          & $\bullet$           & $-$             & \multicolumn{1}{c|}{$\bullet$}     & \multicolumn{1}{c|}{mix}                 & $\bullet$                 & \multicolumn{1}{c|}{$\circ$}                    \\
\textit{Display Multiple Faces}   & MF               & \multicolumn{1}{c|}{$-$}                & $\circ$          & $-$                 & $\bullet$             & \multicolumn{1}{c|}{$\bullet$}     & \multicolumn{1}{c|}{mix}           & $\bullet$                 & \multicolumn{1}{c|}{$-$}                    \\ \hline\hline
\multicolumn{10}{l}{\textit{Audio-based Tasks}} \\ \hline
\textit{Clear   Throat}           & CT               & \multicolumn{1}{c|}{$\bullet$}          & $\bullet$        & $\circ$           & $\bullet$           & \multicolumn{1}{c|}{$\bullet$}     & \multicolumn{1}{c|}{}                  & $\bullet$                 & \multicolumn{1}{c|}{$\circ$}                      \\
\textit{Hold   Musical Note}      & HN               & \multicolumn{1}{c|}{$\bullet$}          & $\circ$          & $\circ$           & $\bullet$           & \multicolumn{1}{c|}{$\bullet$}     & \multicolumn{1}{c|}{}                  & $\bullet$                 & \multicolumn{1}{c|}{$\bullet$}                    \\
\textit{Hum   Tune}               & HT               & \multicolumn{1}{c|}{$\bullet$}          & $\bullet$        & $\bullet$         & $\bullet$           & \multicolumn{1}{c|}{$\bullet$}     & \multicolumn{1}{c|}{}                  & $\bullet$                 & \multicolumn{1}{c|}{$\bullet$}                    \\
\textit{Laugh}                    & L                & \multicolumn{1}{c|}{$\circ$}            & $\bullet$        & $\bullet$         & $\bullet$           & \multicolumn{1}{c|}{$\bullet$}     & \multicolumn{1}{c|}{}                  & $\bullet$                 & \multicolumn{1}{c|}{$\bullet$}                    \\
\textit{Mimic   Speaking Style}   & MS               & \multicolumn{1}{c|}{$\circ$}            & $\bullet$        & $\bullet$         & $\circ$             & \multicolumn{1}{c|}{$\bullet$}     & \multicolumn{1}{c|}{}                  & $\circ$                   & \multicolumn{1}{c|}{$\circ$}                      \\
\textit{Repeat   Accent}          & R                & \multicolumn{1}{c|}{$\circ$}            & $\bullet$        & $\bullet$         & $\circ$             & \multicolumn{1}{c|}{$\bullet$}     & \multicolumn{1}{c|}{}                  & $\circ$                   & \multicolumn{1}{c|}{$\circ$}                      \\
\textit{Sing}                     & S                & \multicolumn{1}{c|}{$\bullet$}          & $\bullet$        & $\bullet$         & $\bullet$           & \multicolumn{1}{c|}{$\bullet$}     & \multicolumn{1}{c|}{}                  & $\bullet$                 & \multicolumn{1}{c|}{$\bullet$}                    \\
\textit{Speak   with Emotion}     & SE               & \multicolumn{1}{c|}{$\bullet$}          & $\bullet$        & $\bullet$         & $\circ$             & \multicolumn{1}{c|}{$\bullet$}     & \multicolumn{1}{c|}{}                  & $\bullet$                 & \multicolumn{1}{c|}{$\bullet$}                    \\
\textit{Yawn}                     & Y                & \multicolumn{1}{c|}{$\circ$}            & $\bullet$        & $\circ$           & $\bullet$           & \multicolumn{1}{c|}{$\bullet$}     & \multicolumn{1}{c|}{}                  & $\bullet$                 & \multicolumn{1}{c|}{$\bullet$}                    \\
\textit{Blow   Noises}            & BN               & \multicolumn{1}{c|}{$\bullet$}          & $\bullet$        & $-$               & $\bullet$           & \multicolumn{1}{c|}{$\bullet$}     & \multicolumn{1}{c|}{bypass}            & $\bullet$                 & \multicolumn{1}{c|}{$-$}                          \\
\textit{Blow   on Mic}            & BM               & \multicolumn{1}{c|}{$\circ$}            & $\bullet$        & $-$               & $\bullet$           & \multicolumn{1}{c|}{$\bullet$}     & \multicolumn{1}{c|}{bypass}            & $\bullet$                 & \multicolumn{1}{c|}{$-$}                          \\
\textit{Clap}                     & Cl               & \multicolumn{1}{c|}{$\bullet$}          & $\circ$          & $-$               & $\bullet$           & \multicolumn{1}{c|}{$\bullet$}     & \multicolumn{1}{c|}{bypass}            & $\bullet$                 & \multicolumn{1}{c|}{$-$}                          \\
\textit{Click   Tongue}           & Clk              & \multicolumn{1}{c|}{$\bullet$}          & $\bullet$        & $-$               & $\bullet$           & \multicolumn{1}{c|}{$\bullet$}     & \multicolumn{1}{c|}{bypass}            & $\bullet$                 & \multicolumn{1}{c|}{$-$}                          \\
\textit{Cough}                    & Co               & \multicolumn{1}{c|}{$\bullet$}          & $\bullet$        & $-$               & $\bullet$           & \multicolumn{1}{c|}{$\bullet$}     & \multicolumn{1}{c|}{bypass}            & $\bullet$                 & \multicolumn{1}{c|}{$-$}                          \\
\textit{Horse   Lips}             & HL               & \multicolumn{1}{c|}{$\circ$}            & $\bullet$        & $-$               & $\bullet$           & \multicolumn{1}{c|}{$\bullet$}     & \multicolumn{1}{c|}{bypass}            & $\bullet$                 & \multicolumn{1}{c|}{$-$}                          \\
\textit{Knock}                    & K                & \multicolumn{1}{c|}{$\circ$}            & $\circ$          & $-$               & $\bullet$           & \multicolumn{1}{c|}{$\bullet$}     & \multicolumn{1}{c|}{bypass}            & $\bullet$                 & \multicolumn{1}{c|}{$-$}                          \\
\textit{Playback   Audio}         & PA               & \multicolumn{1}{c|}{$-$}                & $\bullet$        & $-$               & $\bullet$           & \multicolumn{1}{c|}{$\bullet$}     & \multicolumn{1}{c|}{bypass}            & $\bullet$                 & \multicolumn{1}{c|}{$-$}                          \\
\textit{Raspberry}                & R                & \multicolumn{1}{c|}{$\bullet$}          & $\bullet$        & $-$               & $\bullet$           & \multicolumn{1}{c|}{$\bullet$}     & \multicolumn{1}{c|}{bypass}            & $\bullet$                 & \multicolumn{1}{c|}{$-$}                          \\
\textit{Sound   Effect}           & SFX              & \multicolumn{1}{c|}{$\bullet$}          & $\bullet$        & $-$               & $\bullet$           & \multicolumn{1}{c|}{$\bullet$}     & \multicolumn{1}{c|}{bypass}            & $\bullet$                 & \multicolumn{1}{c|}{$-$}                          \\
\textit{Touch   Mic}              & TM               & \multicolumn{1}{c|}{$\circ$}            & $\bullet$        & $-$               & $\bullet$           & \multicolumn{1}{c|}{$\bullet$}     & \multicolumn{1}{c|}{bypass}            & $\bullet$                 & \multicolumn{1}{c|}{$-$}                          \\
\textit{Type}                     & T                & \multicolumn{1}{c|}{$\circ$}            & $\bullet$        & $-$               & $\bullet$           & \multicolumn{1}{c|}{$\bullet$}     & \multicolumn{1}{c|}{bypass}            & $\bullet$                 & \multicolumn{1}{c|}{$-$}                          \\
\textit{Whistle}                  & W                & \multicolumn{1}{c|}{$-$}                & $\bullet$        & $-$               & $\bullet$           & \multicolumn{1}{c|}{$\bullet$}     & \multicolumn{1}{c|}{bypass}            & $\bullet$                 & \multicolumn{1}{c|}{$-$}                          \\
\textit{Talk   \& Clap}           & T\&C             & \multicolumn{1}{c|}{$\circ$}            & $\bullet$        & $\bullet$         & $\bullet$           & \multicolumn{1}{c|}{$\bullet$}     & \multicolumn{1}{c|}{mix}            & $\bullet$                 & \multicolumn{1}{c|}{$-$}                          \\
\textit{Talk   \& Knock}          & T\&K             & \multicolumn{1}{c|}{$\circ$}            & $\bullet$        & $\bullet$         & $\bullet$           & \multicolumn{1}{c|}{$\bullet$}     & \multicolumn{1}{c|}{mix}            & $\bullet$                 & \multicolumn{1}{c|}{$-$}                          \\
\textit{Talk   \& Playback}       & P                & \multicolumn{1}{c|}{$-$}                & $\bullet$        & $\bullet$         & $\bullet$           & \multicolumn{1}{c|}{$\bullet$}     & \multicolumn{1}{c|}{mix}            & $\bullet$                 & \multicolumn{1}{c|}{$-$}                          \\
\textit{Talk   with Tones}        & TT               & \multicolumn{1}{c|}{$\bullet$}          & $\bullet$        & $\bullet$         & $\bullet$           & \multicolumn{1}{c|}{$\bullet$}     & \multicolumn{1}{c|}{mix}            & $\bullet$                 & \multicolumn{1}{c|}{$\bullet$}                    \\
\textit{Vary   Speed}             & VS               & \multicolumn{1}{c|}{$\bullet$}          & $\bullet$        & $\bullet$         & $\circ$             & \multicolumn{1}{c|}{$\bullet$}     & \multicolumn{1}{c|}{mix}            & $\bullet$                 & \multicolumn{1}{c|}{$\bullet$}                    \\
\textit{Vary   Volume}            & V                & \multicolumn{1}{c|}{$\bullet$}          & $\bullet$        & $\bullet$         & $\circ$             & \multicolumn{1}{c|}{$\bullet$}     & \multicolumn{1}{c|}{mix}            & $\bullet$                 & \multicolumn{1}{c|}{$\bullet$}                    \\ \hline\hline
\multicolumn{10}{l}{$\bullet$: high, $\circ$: medium, $-$: low}
\end{tabular}%
}
\end{table*}

%% file: eval_table.tex
% \usepackage{graphicx}
\begin{table*}[]
\caption{The AUC and EER of voice anomaly detectors/classifiers when used to detect deepfake voices (baseline) and when boosted using our challenges as $\mathcal{R}$. Bold values indicate a case where the challenge improves the detectors ability to spot a deepfake.}
\label{tab:eval}
\centering
\resizebox{0.7\textwidth}{!}{%
\centering
\def\arraystretch{1.15}%  1 is the default, change whatever you need

\begin{tabular}{rcccccccccc}
\multicolumn{1}{c|}{Model}                   & \multicolumn{1}{c|}{Baseline} & R                    & T\&C                 & SE                   & P                    & VS                   & V                    & S                    & HT                   & \multicolumn{1}{c|}{Co}             \\ \hline\hline
\multicolumn{1}{r|}{\textit{SpecRNet}}     & \multicolumn{1}{c|}{0.952}    & 0.914                & 0.538                & 0.796                & 0.825                & 0.922                & 0.92                 & 0.834                & 0.701                & \multicolumn{1}{c|}{0.789}          \\
\multicolumn{1}{r|}{\textit{One-Class}}    & \multicolumn{1}{c|}{0.939}    & \textbf{0.952}       & \textbf{0.967}       & \textbf{0.941}       & \textbf{0.954}       & \textbf{0.958}       & \textbf{0.957}       & \textbf{0.948}       & 0.896                & \multicolumn{1}{c|}{0.832}          \\
\multicolumn{1}{r|}{\textit{GMM-AsvSpoof}} & \multicolumn{1}{c|}{0.949}    & \textbf{0.951}       & \textbf{0.978}       & \textbf{0.953}       & \textbf{0.97}        & \textbf{0.957}       & \textbf{0.949}       & 0.928                & \textbf{0.949}       & \multicolumn{1}{c|}{0.833}          \\
\multicolumn{1}{r|}{\textit{PC-DARTS}}     & \multicolumn{1}{c|}{0.551}    & \textbf{0.568}       & \textbf{0.557}       & \textbf{0.611}       & 0.507                & \textbf{0.586}       & \textbf{0.579}       & \textbf{0.655}       & \textbf{0.675}       & \multicolumn{1}{c|}{\textbf{0.635}} \\
\multicolumn{1}{r|}{\textit{LOF}}          & \multicolumn{1}{c|}{0.678}    & 0.614                & \textbf{0.93}        & 0.635                & \textbf{0.756}       & \textbf{0.771}       & \textbf{0.824}       & 0.593                & \textbf{0.681}       & \multicolumn{1}{c|}{\textbf{0.982}} \\ \hline\hline
                                           & \multicolumn{1}{l}{}          & \multicolumn{1}{l}{} & \multicolumn{1}{l}{} & \multicolumn{1}{l}{} & \multicolumn{1}{l}{} & \multicolumn{1}{l}{} & \multicolumn{1}{l}{} & \multicolumn{1}{l}{} & \multicolumn{1}{l}{} & \multicolumn{1}{l}{}                \\
\multicolumn{1}{c|}{Model}                   & \multicolumn{1}{c|}{Baseline} & R                    & T\&C                 & SE                   & P                    & VS                   & V                    & S                    & HT                   & \multicolumn{1}{c|}{Co}             \\ \hline\hline
\multicolumn{1}{r|}{\textit{SpecRNet}}     & \multicolumn{1}{c|}{0.116}    & 0.163                & 0.475                & 0.285                & 0.261                & 0.155                & 0.154                & 0.245                & 0.354                & \multicolumn{1}{c|}{0.281}          \\
\multicolumn{1}{r|}{\textit{One-Class}}    & \multicolumn{1}{c|}{0.128}    & \textbf{0.123}       & \textbf{0.099}       & 0.133                & \textbf{0.118}       & \textbf{0.112}       & \textbf{0.104}       & \textbf{0.128}       & 0.187                & \multicolumn{1}{c|}{0.259}          \\
\multicolumn{1}{r|}{\textit{GMM-AsvSpoof}} & \multicolumn{1}{c|}{0.122}    & \textbf{0.1}         & \textbf{0.071}       & \textbf{0.099}       & \textbf{0.09}        & \textbf{0.092}       & \textbf{0.115}       & 0.143                & 0.131                & \multicolumn{1}{c|}{0.255}          \\
\multicolumn{1}{r|}{\textit{PC-DARTS}}     & \multicolumn{1}{c|}{0.449}    & \textbf{0.418}       & 0.494                & \textbf{0.386}       & 0.494                & \textbf{0.43}        & \textbf{0.437}       & \textbf{0.366}       & \textbf{0.334}       & \multicolumn{1}{c|}{\textbf{0.415}} \\
\multicolumn{1}{r|}{LOF}                   & \multicolumn{1}{c|}{0.326}    & 0.419                & \textbf{0.122}       & 0.412                & \textbf{0.262}       & \textbf{0.301}       & \textbf{0.26}        & 0.38                 & 0.382                & \multicolumn{1}{c|}{\textbf{0.051}} \\ \hline\hline
\end{tabular}%
}
\end{table*}

%% file: eval_table_mean_eer.tex
% \usepackage{graphicx}

\begin{table*}[t]
\caption{AUC and EER of the video anomaly classifiers/detectors when used as deepfake detectors (baseline) and when boosted using our proposed challenges as $\mathcal{R}$. For AUC, bold values indicate improvement over the baseline; for EER, bold values indicate reduction relative to the baseline.}
\label{tab:auc_eer_video}
\centering
\setlength{\tabcolsep}{3pt} % reduce inter-column white space
\renewcommand{\arraystretch}{1.15}
\resizebox{\textwidth}{!}{%
\begin{tabular}{r|ccccccccccc|ccccccccccc}
\multicolumn{1}{c|}{\multirow{2}{*}{Model}}
& \multicolumn{11}{c|}{AUC} 
& \multicolumn{11}{c}{EER} \\
\cline{2-23}
& Baseline & PC & FC & EM & HO & KS & EM & PCs & SG & TH & SM
& Baseline & PC & FC & EM & HO & KS & EM & PCs & SG & TH & SM \\
\hline\hline

\textit{Capsule-net}
& 0.555 & \textbf{0.677} & \textbf{0.714} & \textbf{0.682} & \textbf{0.771} & \textbf{0.799} & \textbf{0.830} & \textbf{0.846} & \textbf{0.763} & \textbf{0.758} & \textbf{0.683}
& 0.35 & 0.37 & 0.36 & 0.45 & 0.38 & \textbf{0.33} & \textbf{0.27} & \textbf{0.25} & \textbf{0.29} & \textbf{0.30} & 0.41 \\

\textit{CORE}
& 0.777 & \textbf{0.861} & \textbf{0.863} & \textbf{0.850} & \textbf{0.904} & \textbf{0.962} & \textbf{0.987} & \textbf{0.903} & \textbf{0.928} & \textbf{0.911} & \textbf{0.951}
& 0.40 & \textbf{0.26} & \textbf{0.23} & \textbf{0.24} & \textbf{0.19} & \textbf{0.11} & \textbf{0.04} & \textbf{0.12} & \textbf{0.18} & \textbf{0.15} & \textbf{0.11} \\

\textit{F3NET}
& 0.750 & \textbf{0.803} & \textbf{0.843} & \textbf{0.874} & \textbf{0.902} & \textbf{0.908} & \textbf{0.945} & \textbf{0.805} & \textbf{0.830} & \textbf{0.862} & \textbf{0.923}
& 0.35 & 0.40 & \textbf{0.22} & \textbf{0.19} & \textbf{0.21} & \textbf{0.20} & \textbf{0.13} & \textbf{0.21} & \textbf{0.26} & \textbf{0.19} & \textbf{0.23} \\

\textit{FFD}
& 0.688 & \textbf{0.872} & \textbf{0.816} & \textbf{0.851} & \textbf{0.931} & \textbf{0.936} & \textbf{0.936} & \textbf{0.854} & \textbf{0.915} & \textbf{0.886} & \textbf{0.927}
& 0.35 & \textbf{0.18} & \textbf{0.32} & \textbf{0.25} & \textbf{0.15} & \textbf{0.15} & \textbf{0.19} & \textbf{0.26} & \textbf{0.00} & \textbf{0.12} & \textbf{0.18} \\

\textit{RECCE}
& 0.761 & \textbf{0.828} & \textbf{0.809} & \textbf{0.866} & \textbf{0.891} & \textbf{0.953} & \textbf{0.979} & \textbf{0.843} & 0.756 & \textbf{0.870} & \textbf{0.913}
& 0.35 & \textbf{0.22} & \textbf{0.18} & 0.27 & \textbf{21} & \textbf{0.07} & \textbf{0.07} & \textbf{0.34} & \textbf{0.33} & \textbf{0.18} & \textbf{0.17} \\

\textit{SPSL}
& 0.644 & \textbf{0.830} & \textbf{0.884} & \textbf{0.808} & \textbf{0.919} & \textbf{0.918} & \textbf{0.952} & \textbf{0.925} & \textbf{0.929} & \textbf{0.921} & \textbf{0.905}
& 0.45 & \textbf{0.27} & \textbf{0.24} & \textbf{0.27} & \textbf{0.18} & \textbf{0.22} & \textbf{0.10} & \textbf{0.19} & \textbf{0.13} & \textbf{0.18} & \textbf{0.14} \\

\textit{SRM}
& 0.794 & \textbf{0.926} & \textbf{0.862} & \textbf{0.858} & \textbf{0.974} & \textbf{0.952} & \textbf{0.980} & \textbf{0.944} & \textbf{0.903} & \textbf{0.875} & \textbf{0.920}
& 0.25 & \textbf{0.11} & 0.26 & 0.33 & \textbf{0.09} & \textbf{0.11} & \textbf{0.07} & \textbf{0.07} & \textbf{0.16} & \textbf{0.25} & \textbf{0.15} \\

\textit{UCF}
& 0.694 & 0.677 & 0.628 & \textbf{0.804} & 0.610 & 0.649 & \textbf{0.880} & 0.664 & 0.632 & 0.443 & \textbf{0.794}
& 0.35 & \textbf{0.31} & 0.41 & \textbf{0.28} & 0.45 & \textbf{0.35} & \textbf{0.16} & \textbf{0.32} & 0.45 & 0.58 & \textbf{0.23} \\

\textit{Xception}
& 0.733 & \textbf{0.804} & \textbf{0.850} & \textbf{0.864} & \textbf{0.904} & \textbf{0.948} & \textbf{0.993} & \textbf{0.880} & \textbf{0.907} & \textbf{0.941} & \textbf{0.873}
& 0.35 & \textbf{0.29} & \textbf{0.22} & \textbf{0.19} & \textbf{0.19} & \textbf{0.10} & \textbf{0.03} & \textbf{0.17} & \textbf{0.25} & \textbf{0.13} & \textbf{0.19} \\

\hline\hline
\end{tabular}%
}
\end{table*}

%% file: cas-refs.bib
@article{han2024vall,
  title={VALL-E R: Robust and Efficient Zero-Shot Text-to-Speech Synthesis via Monotonic Alignment},
  author={Han, Bing and Zhou, Long and Liu, Shujie and Chen, Sanyuan and Meng, Lingwei and Qian, Yanming and Liu, Yanqing and Zhao, Sheng and Li, Jinyu and Wei, Furu},
  journal={arXiv preprint arXiv:2406.07855},
  year={2024}
}

@article{mirsky2021creation,
  title={The creation and detection of deepfakes: A survey},
  author={Mirsky, Yisroel and Lee, Wenke},
  journal={ACM Computing Surveys (CSUR)},
  volume={54},
  number={1},
  pages={1--41},
  year={2021},
  publisher={ACM New York, NY, USA}
}

@article{Deepfake15:online,
  title={Deepfake Salvador Dal{\'\i} takes selfies with museum visitors},
  author={Lee, Dami},
  journal={The Verge},
  year={2019}
}

@misc{ShamookS49:online,
author = {BBC},
title = {Shamook: Star Wars effects company ILM hires Mandalorian deepfaker - BBC News},
year = {2021}
}

@misc{Reshapin92:online,
author = {Jai Vijayan},
title = {Reshaping the Threat Landscape: Deepfake Cyberattacks Are Here},
year = {2022}
}

@misc{Deepfake76:online,
author = {Karen Hao},
title = {Deepfake porn is ruining women’s lives. Now the law may finally ban it. | MIT Technology Review},
year = {2021}
}

@article{almutairi2022review,
  title={A Review of Modern Audio Deepfake Detection Methods: Challenges and Future Directions},
  author={Almutairi, Zaynab and Elgibreen, Hebah},
  journal={Algorithms},
  volume={15},
  number={5},
  year={2022},
  publisher={MDPI}
}

@inproceedings{lei2020siamese,
  title={Siamese Convolutional Neural Network Using Gaussian Probability Feature for Spoofing Speech Detection.},
  author={Lei, Zhenchun and Yang, Yingen and Liu, Changhong and Ye, Jihua},
  booktitle={INTERSPEECH},
  pages={1116--1120},
  year={2020}
}

@article{lai2019assert,
  title={ASSERT: Anti-spoofing with squeeze-excitation and residual networks},
  author={Lai, Cheng-I and Chen, Nanxin and Villalba, Jes{\'u}s and Dehak, Najim},
  journal={arXiv preprint arXiv:1904.01120},
  year={2019}
}

@inproceedings{rawnet2,
  title={End-to-end anti-spoofing with rawnet2},
  author={Tak, Hemlata and Patino, Jose and Todisco, Massimiliano and Nautsch, Andreas and Evans, Nicholas and Larcher, Anthony},
  booktitle={ICASSP 2021-2021 IEEE International Conference on Acoustics, Speech and Signal Processing (ICASSP)},
  pages={6369--6373},
  year={2021},
  organization={IEEE}
}

@article{kawa2022specrnet,
  title={SpecRNet: Towards Faster and More Accessible Audio DeepFake Detection},
  author={Kawa, Piotr and Plata, Marcin and Syga, Piotr},
  journal={arXiv preprint},
  year={2022}
}

@article{khochare2022deep,
  title={A deep learning framework for audio deepfake detection},
  author={Khochare, Janavi and Joshi, Chaitali and Yenarkar, Bakul and Suratkar, Shraddha and Kazi, Faruk},
  journal={Arabian Journal for Science and Engineering},
  volume={47},
  number={3},
  pages={3447--3458},
  year={2022},
  publisher={Springer}
}

@inproceedings{khalid2020oc,
  title={OC-FakeDect: Classifying deepfakes using one-class variational autoencoder},
  author={Khalid, Hasam and Woo, Simon S},
  booktitle={Proceedings of the IEEE/CVF conference on computer vision and pattern recognition workshops},
  pages={656--657},
  year={2020}
}

@inproceedings{agarwal2020detecting,
  title={Detecting deep-fake videos from phoneme-viseme mismatches},
  author={Agarwal, Shruti and Farid, Hany and Fried, Ohad and Agrawala, Maneesh},
  booktitle={Proceedings of the IEEE/CVF conference on computer vision and pattern recognition workshops},
  pages={660--661},
  year={2020}
}

@article{borrelli2021synthetic,
  title={Synthetic speech detection through short-term and long-term prediction traces},
  author={Borrelli, Clara and Bestagini, Paolo and Antonacci, Fabio and Sarti, Augusto and Tubaro, Stefano},
  journal={EURASIP Journal on Information Security},
  volume={2021},
  number={1},
  pages={1--14},
  year={2021},
  publisher={Springer}
}

@article{zhang2021one,
  title={One-class learning towards synthetic voice spoofing detection},
  author={Zhang, You and Jiang, Fei and Duan, Zhiyao},
  journal={IEEE Signal Processing Letters},
  volume={28},
  pages={937--941},
  year={2021},
  publisher={IEEE}
}

@inproceedings{camacho2021fake,
  title={Fake speech recognition using deep learning},
  author={Camacho, Steven and Ballesteros, Dora Maria and Renza, Diego},
  booktitle={Workshop on Engineering Applications},
  year={2021},
  organization={Springer}
}

@article{liu2021identification,
  title={Identification of fake stereo audio using SVM and CNN},
  author={Liu, Tianyun and Yan, Diqun and Wang, Rangding and Yan, Nan and Chen, Gang},
  journal={Information},
  volume={12},
  number={7},
  year={2021},
  publisher={MDPI}
}

@article{kawa2022attack,
  title={Attack Agnostic Dataset: Towards Generalization and Stabilization of Audio DeepFake Detection},
  author={Kawa, Piotr and Plata, Marcin and Syga, Piotr},
  journal={arXiv preprint},
  year={2022}
}

@article{muller2022does,
  title={Does Audio Deepfake Detection Generalize?},
  author={M{\"u}ller, Nicolas M and Czempin, Pavel and Dieckmann, Franziska and Froghyar, Adam and B{\"o}ttinger, Konstantin},
  journal={arXiv preprint arXiv:2203.16263},
  year={2022}
}

@article{pianese2022deepfake,
  title={Deepfake audio detection by speaker verification},
  author={Pianese, Alessandro and Cozzolino, Davide and Poggi, Giovanni and Verdoliva, Luisa},
  journal={arXiv preprint},
  year={2022}
}

@article{todisco2019asvspoof,
  title={ASVspoof 2019: Future horizons in spoofed and fake audio detection},
  author={Todisco, Massimiliano and Wang, Xin and Vestman, Ville and Sahidullah, Md and Delgado, H{\'e}ctor and Nautsch, Andreas and Yamagishi, Junichi and Evans, Nicholas and Kinnunen, Tomi and Lee, Kong Aik},
  journal={arXiv preprint arXiv:1904.05441},
  year={2019}
}

@article{arif2021voice,
  title={Voice spoofing countermeasure for logical access attacks detection},
  author={Arif, Tuba and Javed, Ali and Alhameed, Mohammed and Jeribi, Fathe and Tahir, Ali},
  journal={IEEE Access},
  volume={9},
  year={2021},
  publisher={IEEE}
}

@inproceedings{li2021deepfake,
  title={Deepfake detection using robust spatial and temporal features from facial landmarks},
  author={Li, Meng and Liu, Beibei and Hu, Yongjian and Zhang, Liepiao and Wang, Shiqi},
  booktitle={2021 IEEE International Workshop on Biometrics and Forensics (IWBF)},
  pages={1--6},
  year={2021},
  organization={IEEE}
}

@inproceedings{yang2019exposing,
  title={Exposing deep fakes using inconsistent head poses},
  author={Yang, Xin and Li, Yuezun and Lyu, Siwei},
  booktitle={ICASSP 2019-2019 IEEE International Conference on Acoustics, Speech and Signal Processing (ICASSP)},
  pages={8261--8265},
  year={2019},
  organization={IEEE}
}

@inproceedings{amerini2019deepfake,
  title={Deepfake video detection through optical flow based cnn},
  author={Amerini, Irene and Galteri, Leonardo and Caldelli, Roberto and Del Bimbo, Alberto},
  booktitle={Proceedings of the IEEE/CVF international conference on computer vision workshops},
  pages={0--0},
  year={2019}
}

@article{wodajo2021deepfake,
  title={Deepfake video detection using convolutional vision transformer},
  author={Wodajo, Deressa and Atnafu, Solomon},
  journal={arXiv preprint arXiv:2102.11126},
  year={2021}
}

@article{masood2023deepfakes,
  title={Deepfakes generation and detection: State-of-the-art, open challenges, countermeasures, and way forward},
  author={Masood, Momina and Nawaz, Mariam and Malik, Khalid Mahmood and Javed, Ali and Irtaza, Aun and Malik, Hafiz},
  journal={Applied intelligence},
  volume={53},
  number={4},
  pages={3974--4026},
  year={2023},
  publisher={Springer}
}

@article{guera2019we,
  title={We need no pixels: Video manipulation detection using stream descriptors},
  author={G{\"u}era, David and Baireddy, Sriram and Bestagini, Paolo and Tubaro, Stefano and Delp, Edward J},
  journal={arXiv preprint arXiv:1906.08743},
  year={2019}
}

@article{ciftci2020fakecatcher,
  title={Fakecatcher: Detection of synthetic portrait videos using biological signals},
  author={Ciftci, Umur Aybars and Demir, Ilke and Yin, Lijun},
  journal={IEEE transactions on pattern analysis and machine intelligence},
  year={2020},
  publisher={IEEE}
}

@article{jung2020deepvision,
  title={Deepvision: Deepfakes detection using human eye blinking pattern},
  author={Jung, Tackhyun and Kim, Sangwon and Kim, Keecheon},
  journal={IEEE Access},
  volume={8},
  pages={83144--83154},
  year={2020},
  publisher={IEEE}
}

@article{li2018exposing,
  title={Exposing deepfake videos by detecting face warping artifacts},
  author={Li, Yuezun and Lyu, Siwei},
  journal={arXiv preprint arXiv:1811.00656},
  year={2018}
}

@inproceedings{guera2018deepfake,
  title={Deepfake video detection using recurrent neural networks},
  author={G{\"u}era, David and Delp, Edward J},
  booktitle={2018 15th IEEE international conference on advanced video and signal based surveillance (AVSS)},
  pages={1--6},
  year={2018},
  organization={IEEE}
}

@article{li2018ictu,
  title={In ictu oculi: Exposing ai generated fake face videos by detecting eye blinking},
  author={Li, Yuezun and Chang, Ming-Ching and Lyu, Siwei},
  journal={arXiv preprint arXiv:1806.02877},
  year={2018}
}

@inproceedings{montserrat2020deepfakes,
  title={Deepfakes detection with automatic face weighting},
  author={Montserrat, Daniel Mas and Hao, Hanxiang and Yarlagadda, Sri K and Baireddy, Sriram and Shao, Ruiting and Horv{\'a}th, J{\'a}nos and Bartusiak, Emily and Yang, Justin and Guera, David and Zhu, Fengqing and others},
  booktitle={Proceedings of the IEEE/CVF conference on computer vision and pattern recognition workshops},
  pages={668--669},
  year={2020}
}

@article{de2020deepfake,
  title={Deepfake detection using spatiotemporal convolutional networks},
  author={De Lima, Oscar and Franklin, Sean and Basu, Shreshtha and Karwoski, Blake and George, Annet},
  journal={arXiv preprint arXiv:2006.14749},
  year={2020}
}

@inproceedings{fernandes2019predicting,
  title={Predicting heart rate variations of deepfake videos using neural ode},
  author={Fernandes, Steven and Raj, Sunny and Ortiz, Eddy and Vintila, Iustina and Salter, Margaret and Urosevic, Gordana and Jha, Sumit},
  booktitle={Proceedings of the IEEE/CVF international conference on computer vision workshops},
  pages={0--0},
  year={2019}
}

@article{yang2021msta,
  title={MSTA-Net: Forgery detection by generating manipulation trace based on multi-scale self-texture attention},
  author={Yang, Jiachen and Xiao, Shuai and Li, Aiyun and Lu, Wen and Gao, Xinbo and Li, Yang},
  journal={IEEE transactions on circuits and systems for video technology},
  volume={32},
  number={7},
  pages={4854--4866},
  year={2021},
  publisher={IEEE}
}

@inproceedings{afchar2018mesonet,
  title={Mesonet: a compact facial video forgery detection network},
  author={Afchar, Darius and Nozick, Vincent and Yamagishi, Junichi and Echizen, Isao},
  booktitle={2018 IEEE international workshop on information forensics and security (WIFS)},
  pages={1--7},
  year={2018},
  organization={IEEE}
}

@inproceedings{ahn2003captcha,
  title={CAPTCHA: Using hard AI problems for security},
  author={Ahn, Luis von and Blum, Manuel and Hopper, Nicholas J and Langford, John},
  booktitle={International conference on the theory and applications of cryptographic techniques},
  pages={294--311},
  year={2003},
  organization={Springer}
}

@article{yamagishi2021asvspoof,
  title={ASVspoof 2021: accelerating progress in spoofed and deepfake speech detection},
  author={Yamagishi, Junichi and Wang, Xin and Todisco, Massimiliano and Sahidullah, Md and Patino, Jose and Nautsch, Andreas and Liu, Xuechen and Lee, Kong Aik and Kinnunen, Tomi and Evans, Nicholas and others},
  journal={arXiv preprint arXiv:2109.00537},
  year={2021}
}

@article{ge2021raw,
  title={Raw differentiable architecture search for speech deepfake and spoofing detection},
  author={Ge, Wanying and Patino, Jose and Todisco, Massimiliano and Evans, Nicholas},
  journal={arXiv preprint arXiv:2107.12212},
  year={2021}
}

@article{speechbrain,
  title={SpeechBrain: A general-purpose speech toolkit},
  author={Ravanelli, Mirco and Parcollet, Titouan and Plantinga, Peter and Rouhe, Aku and Cornell, Samuele and Lugosch, Loren and Subakan, Cem and Dawalatabad, Nauman and Heba, Abdelwahab and Zhong, Jianyuan and others},
  journal={arXiv preprint arXiv:2106.04624},
  year={2021}
}

@article{perov2020deepfacelab,
  title={DeepFaceLab: Integrated, flexible and extensible face-swapping framework},
  author={Perov, Ivan and Gao, Daiheng and Chervoniy, Nikolay and Liu, Kunlin and Marangonda, Sugasa and Um{\'e}, Chris and Dpfks, Mr and Facenheim, Carl Shift and RP, Luis and Jiang, Jian and others},
  journal={arXiv preprint arXiv:2005.05535},
  year={2020}
}

@inproceedings{DBLP:conf/mm/ChenCNG20,
  author    = {Renwang Chen and
               Xuanhong Chen and
               Bingbing Ni and
               Yanhao Ge},
  title     = {SimSwap: An Efficient Framework For High Fidelity Face Swapping},
  booktitle = {{MM} '20: The 28th {ACM} International Conference on Multimedia},
  year      = {2020}
}

@inproceedings{he2016deep,
  title={Deep residual learning for image recognition},
  author={He, Kaiming and Zhang, Xiangyu and Ren, Shaoqing and Sun, Jian},
  booktitle={Proceedings of the IEEE conference on computer vision and pattern recognition},
  pages={770--778},
  year={2016}
}

@inproceedings{tran2018closer,
  title={A closer look at spatiotemporal convolutions for action recognition},
  author={Tran, Du and Wang, Heng and Torresani, Lorenzo and Ray, Jamie and LeCun, Yann and Paluri, Manohar},
  booktitle={Proceedings of the IEEE conference on Computer Vision and Pattern Recognition},
  pages={6450--6459},
  year={2018}
}

@inproceedings{ni2022core,
  title={Core: Consistent representation learning for face forgery detection},
  author={Ni, Yunsheng and Meng, Depu and Yu, Changqian and Quan, Chengbin and Ren, Dongchun and Zhao, Youjian},
  booktitle={Proceedings of the IEEE/CVF conference on computer vision and pattern recognition},
  pages={12--21},
  year={2022}
}

@inproceedings{qian2020thinking,
  title={Thinking in frequency: Face forgery detection by mining frequency-aware clues},
  author={Qian, Yuyang and Yin, Guojun and Sheng, Lu and Chen, Zixuan and Shao, Jing},
  booktitle={European conference on computer vision},
  pages={86--103},
  year={2020},
  organization={Springer}
}

@inproceedings{yan2023ucf,
  title={Ucf: Uncovering common features for generalizable deepfake detection},
  author={Yan, Zhiyuan and Zhang, Yong and Fan, Yanbo and Wu, Baoyuan},
  booktitle={Proceedings of the IEEE/CVF International Conference on Computer Vision},
  pages={22412--22423},
  year={2023}
}

@inproceedings{nguyen2019capsule,
  title={Capsule-forensics: Using capsule networks to detect forged images and videos},
  author={Nguyen, Huy H and Yamagishi, Junichi and Echizen, Isao},
  booktitle={ICASSP 2019-2019 IEEE international conference on acoustics, speech and signal processing (ICASSP)},
  pages={2307--2311},
  year={2019},
  organization={IEEE}
}

@inproceedings{rossler2019faceforensics++,
  title={Faceforensics++: Learning to detect manipulated facial images},
  author={Rossler, Andreas and Cozzolino, Davide and Verdoliva, Luisa and Riess, Christian and Thies, Justus and Nie{\ss}ner, Matthias},
  booktitle={Proceedings of the IEEE/CVF international conference on computer vision},
  pages={1--11},
  year={2019}
}

@inproceedings{liu2021spatial,
  title={Spatial-phase shallow learning: rethinking face forgery detection in frequency domain},
  author={Liu, Honggu and Li, Xiaodan and Zhou, Wenbo and Chen, Yuefeng and He, Yuan and Xue, Hui and Zhang, Weiming and Yu, Nenghai},
  booktitle={Proceedings of the IEEE/CVF conference on computer vision and pattern recognition},
  pages={772--781},
  year={2021}
}

@inproceedings{luo2021generalizing,
  title={Generalizing face forgery detection with high-frequency features},
  author={Luo, Yuchen and Zhang, Yong and Yan, Junchi and Liu, Wei},
  booktitle={Proceedings of the IEEE/CVF conference on computer vision and pattern recognition},
  pages={16317--16326},
  year={2021}
}

@inproceedings{dang2020detection,
  title={On the detection of digital face manipulation},
  author={Dang, Hao and Liu, Feng and Stehouwer, Joel and Liu, Xiaoming and Jain, Anil K},
  booktitle={Proceedings of the IEEE/CVF Conference on Computer Vision and Pattern recognition},
  pages={5781--5790},
  year={2020}
}

@inproceedings{cao2022end,
  title={End-to-end reconstruction-classification learning for face forgery detection},
  author={Cao, Junyi and Ma, Chao and Yao, Taiping and Chen, Shen and Ding, Shouhong and Yang, Xiaokang},
  booktitle={Proceedings of the IEEE/CVF Conference on Computer Vision and Pattern Recognition},
  pages={4113--4122},
  year={2022}
}

@inproceedings{DeepfakeBench_YAN_NEURIPS2023,
 author = {Yan, Zhiyuan and Zhang, Yong and Yuan, Xinhang and Lyu, Siwei and Wu, Baoyuan},
 booktitle = {Advances in Neural Information Processing Systems},
 editor = {A. Oh and T. Neumann and A. Globerson and K. Saenko and M. Hardt and S. Levine},
 pages = {4534--4565},
 publisher = {Curran Associates, Inc.},
 title = {DeepfakeBench: A Comprehensive Benchmark of Deepfake Detection},
 url = {https://proceedings.neurips.cc/paper_files/paper/2023/file/0e735e4b4f07de483cbe250130992726-Paper-Datasets_and_Benchmarks.pdf},
 volume = {36},
 year = {2023}
}

@misc{facefusion,
author = {FaceFusion},
title = {GitHub - facefusion/facefusion: Industry leading face manipulation platform},
howpublished = {\url{https://github.com/facefusion/facefusion}},
month = {},
year = {2024},
note = {(Accessed on 16/04/2026)}
}

@misc{roop,
author = {s0mdv3},
title = {GitHub - s0md3v/roop: one-click face swap},
howpublished = {\url{https://github.com/s0md3v/roop}},
month = {},
year = {2024},
note = {(Accessed on 16/04/2026)}
}

@misc{VoiceClo10:online,
author = {ReSpeecher},
title = {Voice Cloning Software for Content Creators | Respeecher},
howpublished = {\url{https://www.respeecher.com/}},
}

@misc{Deepfake46:online,
author = {Jane Wakefield},
title = {Deepfake presidents used in Russia-Ukraine war - BBC News},
howpublished = {\url{https://www.bbc.com/news/technology-60780142}},
month = {March},
year = {2022},
note = {(Accessed on 16/04/2026)}
}

@incollection{NIPS2019_8935,
	title = {First Order Motion Model for Image Animation},
	author = {Siarohin, Aliaksandr and Lathuili\`{e}re, St\'{e}phane and Tulyakov, Sergey and Ricci, Elisa and Sebe, Nicu},
	booktitle = {Advances in Neural Information Processing Systems 32},
	pages = {7135--7145},
	year = {2019},
	publisher = {Curran Associates, Inc.}
}

@misc{AdaIN,
  doi = {10.48550/ARXIV.1904.05742},
  
  url = {https://arxiv.org/abs/1904.05742},
  
  author = {Chou, Ju-chieh and Yeh, Cheng-chieh and Lee, Hung-yi},
  
  title = {One-shot Voice Conversion by Separating Speaker and Content Representations with Instance Normalization},
  
  publisher = {arXiv},
  
  year = {2019},
  
  copyright = {arXiv.org perpetual, non-exclusive license}
}

@misc{FragmentVC,
  doi = {10.48550/ARXIV.2010.14150},
  
  url = {https://arxiv.org/abs/2010.14150},
  
  author = {Lin, Yist Y. and Chien, Chung-Ming and Lin, Jheng-Hao and Lee, Hung-yi and Lee, Lin-shan},
  
  title = {FragmentVC: Any-to-Any Voice Conversion by End-to-End Extracting and Fusing Fine-Grained Voice Fragments With Attention},
  
  publisher = {arXiv},
  
  year = {2020},
  
  copyright = {arXiv.org perpetual, non-exclusive license}
}

@misc{Fraudste87:online,
author = {Catherine Stupp},
title = {Fraudsters Used AI to Mimic CEO’s Voice in Unusual Cybercrime Case - WSJ},
month = {August},
year = {2019}}

@misc{European87:online,
author = {Andrew Roth },
title = {European MPs targeted by deepfake video calls imitating Russian opposition | Russia | The Guardian},
year = {2021}
}

@misc{Fraudste98:online,
author = {Thomas Brewster},
title = {Fraudsters Cloned Company Director’s Voice In \$35 Million Bank Heist, Police Find},
year = {2021}
}

@misc{Binancee98:online,
author = {James Vincent},
title = {Binance executive claims scammers made a deepfake of him - The Verge},
year = {2022}
}

@misc{Assem-VC,
  doi = {10.48550/ARXIV.2104.00931},
  
  url = {https://arxiv.org/abs/2104.00931},
  
  author = {Kim, Kang-wook and Park, Seung-won and Lee, Junhyeok and Joe, Myun-chul},
  
  title = {Assem-VC: Realistic Voice Conversion by Assembling Modern Speech Synthesis Techniques},
  
  publisher = {arXiv},
  
  year = {2021},
  
  copyright = {arXiv.org perpetual, non-exclusive license}
}

@inproceedings{carlini2020evading,
  title={Evading deepfake-image detectors with white-and black-box attacks},
  author={Carlini, Nicholas and Farid, Hany},
  booktitle={Proceedings of the IEEE/CVF conference on computer vision and pattern recognition workshops},
  pages={658--659},
  year={2020}
}

@misc{MediumVC,
  doi = {10.48550/ARXIV.2110.02500},
  
  url = {https://arxiv.org/abs/2110.02500},
  
  author = {Gu, Yewei and Zhang, Zhenyu and Yi, Xiaowei and Zhao, Xianfeng},
  
  title = {MediumVC: Any-to-any voice conversion using synthetic specific-speaker speeches as intermedium features},
  
  publisher = {arXiv},
  
  year = {2021},
  
  copyright = {Creative Commons Attribution 4.0 International}
}

@article{StarGANv2-VC,
  author    = {Yinghao Aaron Li and
               Ali Zare and
               Nima Mesgarani},
  title     = {StarGANv2-VC: {A} Diverse, Unsupervised, Non-parallel Framework for
               Natural-Sounding Voice Conversion},
  journal   = {CoRR},
  volume    = {abs/2107.10394},
  year      = {2021},
  url       = {https://arxiv.org/abs/2107.10394},
  eprinttype = {arXiv},
  eprint    = {2107.10394},
  bibsource = {dblp computer science bibliography, https://dblp.org}
}

@misc{Internet56:online,
author = {FBI},
title = {Internet Crime Complaint Center (IC3) | Deepfakes and Stolen PII Utilized to Apply for Remote Work Positions},
howpublished = {\url{https://www.ic3.gov/Media/Y2022/PSA220628}},
month = {June},
year = {2022}
}

@article{phone_filter,
  title={Processing the telephone speech signal for the hearing impaired},
  author={Terry, Mark and Bright, Kathryn and Durian, Mike and Kepler, Laura and Sweetman, Richard and Grim, Michael},
  journal={Ear and Hearing},
  volume={13},
  number={2},
  pages={70--79},
  year={1992},
  publisher={LWW}
}

@misc{Indiasge48:online,
author = {Sophie Landrin},
title = {India's general election is being impacted by deepfakes},
howpublished = {\url{https://www.lemonde.fr/en/pixels/article/2024/05/21/india-s-general-election-is-being-impacted-by-deepfakes_6672168_13.html?utm_source=chatgpt.com#}},
month = {},
year = {2024},
note = {(Accessed on 16/04/2026)}
}

@misc{Melonisu84:online,
author = {Tom Kington},
title = {Meloni sues deepfake porn creator ‘to protect women’},
howpublished = {https://www.thetimes.com/world/europe/article/meloni-sues-deepfake-porn-creator-to-protect-women-rxvjz08x7},
month = {},
year = {2024},
note = {(Accessed on 16/04/2026)}
}

@misc{Deepfake67:online,
author = {Jessica Bahr},
title = {Deepfakes, blackmail and heartbreak: This is what sextortion looks like | SBS News},
howpublished = {https://www.sbs.com.au/news/article/deepfakes-blackmail-and-heartbreak-this-is-what-sextortion-looks-like/rejaipy9p},
month = {},
year = {2024},
note = {(Accessed on 16/04/2026)}
}

@misc{Millions20:online,
author = {James Halpin},
title = {Millions of Brits warned over AI 'video calls' duping employees into £20m deepfake scam | The Irish Sun},
howpublished = {\url{https://www.thesun.ie/tech/12989965/ai-deepfake-scam-arup-hong-kong-video/?utm_source=chatgpt.com}},
month = {},
year = {2024},
note = {(Accessed on 16/04/2026)}
}

@misc{Sophisti0:online,
author = {Dan Merica},
title = {Sophistication of AI-backed operation targeting senator points to future of deepfake schemes | AP News},
howpublished = {https://apnews.com/article/deepfake-cardin-ai-artificial-intelligence-879a6c2ca816c71d9af52a101dedb7ff},
month = {},
year = {2024},
note = {(Accessed on 16/04/2026)}
}

@misc{USmother64:online,
author = {Erum Salam},
title = {US mother gets call from ‘kidnapped daughter’ – but it’s really an AI scam | Arizona | The Guardian},
howpublished = {\url{https://www.theguardian.com/us-news/2023/jun/14/ai-kidnapping-scam-senate-hearing-jennifer-destefano}},
month = {},
year = {2023},
note = {(Accessed on 16/04/2026)}
}

@misc{Binancee50:online,
author = {Luke Hurst},
title = {Binance executive says scammers created deepfake ‘hologram’ of him to trick crypto developers | Euronews},
howpublished = {https://www.euronews.com/next/2022/08/24/binance-executive-says-scammers-created-deepfake-hologram-of-him-to-trick-crypto-developer},
month = {},
year = {2022},
note = {(Accessed on 16/04/2026)}
}

@inproceedings{yasur2023deepfake,
  title={Deepfake captcha: a method for preventing fake calls},
  author={Yasur, Lior and Frankovits, Guy and Grabovski, Fred M and Mirsky, Yisroel},
  booktitle={Proceedings of the 2023 ACM Asia Conference on Computer and Communications Security},
  pages={608--622},
  year={2023}
}
